\documentclass[aps, english, pra, twocolumn, superscriptaddress, nofootinbib, longbibliography]{revtex4-2}

\usepackage[english]{babel}
\usepackage{tikz}
\usepackage{amsmath}
\usetikzlibrary{quantikz2}
\usepackage{multirow}

\usepackage{xcolor}
\usepackage{amsmath}
\usepackage{graphicx}
\usepackage[colorlinks=true, allcolors=blue]{hyperref}
\def\vector#1.{\begin{pmatrix} #1 \end{pmatrix}}

\begin{document}
\title{Pauli Cloners for Pauli Channels}

\date{January 31, 2026}
\author{S. F. Kerstan}
\author{M. Gallezot}
\author{T. Decker}
\author{M. Braun}
\author{N. Hegemann}
\affiliation{JoS QUANTUM GmbH, c/o Tech Quartier, Platz der Einheit 2, \protect 60327 Frankfurt am Main, Germany \\ \{\emph{sven.kerstan, marcelin.gallezot, thomas.decker}\}@jos-quantum.de}

\begin{abstract}
We present a quantum circuit architecture for the one-to-two cloning of $N$-qubit registers. It implements the broad class of Pauli cloners by extending the Niu--Griffiths architecture to multi-qubit systems. In the single-qubit case, we provide explicit constructions for asymmetric universal, phase covariant and biased cloners. We explore the fundamental relationship between Pauli errors, mutually unbiased bases and Pauli cloning. Furthermore, we demonstrate how Pauli cloners can be tailored to specific noise models in the context of quantum communication, especially quantum key distribution.
\end{abstract}

\maketitle

\section*{Introduction}
More than 25 years ago, two distinct quantum circuits were suggested for the one-to-two cloning of qubits: the Niu--Griffiths cloner (NG cloner)~\cite{niugriffiths} and the Quantum Information Distributor (QID)~\cite{qid1997}. Both were motivated, at least in part, by developments in quantum cryptography, particularly Quantum Key Distribution (QKD; for a review, see e.g.~\cite{QKDreview}). Indeed, both circuits can implement the Universal Quantum Cloning Machine (UQCM), which was the key to optimal individual attacks on the six-state QKD protocol~\cite{sixstate}.\\
Another shared feature of these approaches is that they can be split up into a \textit{software} or \textit{program} part -- the preparation of a quantum state on the ancillary qubits -- and a \textit{hardware} part, consisting of a fixed set of gates that perform the cloning according to the program. Furthermore, in the single-qubit case, both circuits implement Pauli cloners; that is, the disturbance they introduce into the clones is Pauli noise.\\
The field evolved quickly over the next few years. While early work focused on symmetric cloners (where both copies share the same fidelity), the scope soon extended to asymmetric versions of the UQCM~\cite{cerf1998}. Subsequently, the QID was shown to generalize the UQCM to arbitrary dimensional Hilbert spaces~\cite{QID}. Beyond the universal case, the Phase Covariant Cloning Machine (PCCM)~\cite{PCCM}—the optimal attack on the BB84 protocol~\cite{BB84}—was implemented using the QID in~\cite{PCCMinQID}. For a historical review and comprehensive overview of quantum cloning, we refer the reader to~\cite{cerfreview, cloningreview1, cloningreviewfan}.\\
However, although it was recognized early on that these circuits could realize many more cloning machines~\cite{niugriffiths, qid1997}, concrete results regarding the implementation and application of all possible cloning machines remain, to our knowledge, largely unexplored. This is likely because state-dependent cloners have received relatively little attention (for a notable exception, see~\cite{stateDep}). Yet, many natural cloning problems for Pauli cloners are, as we will demonstrate, state dependent. A property of our Pauli cloners is that they clone all basis states belonging to each mutually unbiased basis (MUB; see e.g.~\cite{MUBs}) with the same fidelity, while the fidelity may differ between two different MUBs. Thus, Pauli cloners are inherently \textit{biased cloners}. The special case of no bias corresponds to the well-known UQCM. Note that maximizing the cloning quality on only a subset of MUBs, as the PCCM does, is a specific instance of such bias.\\
Our aim in this work is twofold. First, we illustrate the role of "all the other" cloning machines hiding in the original circuits, providing circuit implementations motivated by practical applications. Second, we generalize these concepts from single-qubit to $N$-qubit systems. We show that the natural path to this generalization is a straightforward extension of the Niu--Griffiths cloner~\cite{niugriffiths}, which we introduce in Section~\ref{paulichannels}. Readers familiar with cloning machines may object that the QID already provides a generalization to arbitrary dimensions. However, as noted in~\cite{cerfAsymmetric}, the higher dimensional QID does not implement Pauli cloners. While the QID implements Pauli cloners for single qubits (see Appendix~\ref{QIDsinglequbit}), it is significantly outperformed by our $N$-qubit Pauli cloners for Pauli channels with $N > 1$. We provide comparative examples in Section~\ref{twoqubitcloners}.\\
Our results have potential applications on the security and performance of present and future QKD and quantum communication applications, as well as one of the founding ideas in the field of quantum information: quantum money~\cite{wiesner1983}.\\
Finally, we note that this work extensively utilizes the Quantum Machine Learning (QML) ideas and tools developed in~\cite{qkdasQML}. Appendix~\ref{appendix:qcl} provides background on the application of these techniques to our results.

\section{Single-qubit one-to-two cloning with the Niu--Griffiths cloner} \label{1qubitNG}
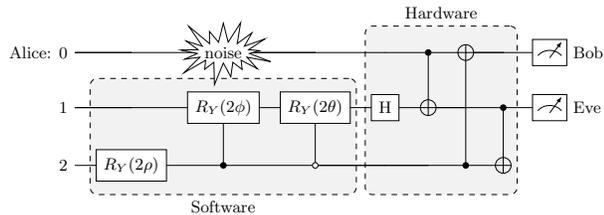
\begin{figure}[h!]
\centering
\begin{tikzpicture}
\node[scale=0.7]{
\tikzset{
noisy/.style={starburst,fill=white,draw=black,line
width=0.5pt,inner xsep=-4pt,inner ysep=-3pt}
}
\begin{quantikz}[thin lines, classical gap=0.09cm, column sep = 0.4cm, color=black,background color=white]
\lstick{Alice: 0}&\qw &\gate[1,style={noisy},label style=black]{\text{noise}} &\qw & \gategroup[3,steps=4,style={dashed,rounded corners,fill=black!5, inner xsep=0pt},background]{Hardware} &\ctrl{1} &\targ{}&\qw&\meter{} \rstick{Bob} \\
\lstick{1}&\qw \gategroup[2,steps=3,style={dashed,rounded corners,fill=black!5, inner xsep=0pt}, background, label style={label
position=below,anchor=north,yshift=-0.2cm}]{Software} &\gate{R_Y(2 \phi)}&\gate{R_Y(2 \theta)} & \gate{\text{H}} &\targ{}&\qw &\ctrl{1} &\meter{} \rstick{Eve} \\
\lstick{2}&\gate{R_Y(2 \rho)}&\ctrl{-1}&\ctrl[open]{-1}&\qw &\qw &\ctrl{-2}&\targ{}
\end{quantikz}
};
\end{tikzpicture}
    \caption{Quantum circuit implementation of the single-qubit Niu--Griffiths cloner. Qubit 0 serves as both input and first output, while qubit 1 is the second output. Qubit 0 can be affected by Pauli noise, but only until Eve's controlled gates begin to act. The program state is initialized with three $R_Y$-rotations, allowing only real states parametrized by the three angles $\rho$, $\phi$ and $\theta$. The hardware part is implemented with an H gate and three CNOTs.}
    \label{1qubitNGcircuit}
\end{figure}

We start with single-qubit cloners. Such devices serve as a typical building block for eavesdropping attack on a quantum channel, such as those used for QKD between two parties Alice and Bob. In this scenario, the eavesdropper, Eve, inserts a cloner into the channel to obtain clones of the transmitted qubits. Figure~\ref{1qubitNGcircuit} illustrates a quantum circuit implementation of such a setup using the Niu--Griffiths cloner. One motivation for the design of this cloner as well as the QID was to implement optimal eavesdropping attacks on the six-state QKD protocol~\cite{sixstate}. This optimal cloner is often called universal quantum cloning machine (UQCM). The software states that program the Niu--Griffiths cloner to implement the UQCM were given in~\cite{niugriffiths}. Other interesting cloners can be implemented using real software states, we justify later why only real states need to be considered.\\

In this section, we provide explicit circuit parameterizations for some of the well known single-qubit quantum cloning machines. These include the asymmetric UQCM, the asymmetric PCCM, which implements the optimal attack for the BB84 QKD protocol~\cite{BB84}, and asymmetric imbalanced cloners~\cite{qkdasQML}, which we conjecture to implement optimal attacks on the BB84 protocol in the presence of Pauli noise.

\subsection{Niu--Griffiths cloners are Pauli cloners}
Before we proceed, we would like to point out that the Niu-Griffith cloner is a Pauli cloner. This term, used by Cerf in~\cite{cerfAsymmetric} to characterize the single-qubit QID, implies that it introduces Pauli noise in both output qubits. An inspection of Figure~\ref{1qubitNGcircuit} confirms that this is true for the NG cloner. The qubit that Alice sends to Bob (qubit 0) is affected by two gate operations. The first is a CNOT gate on qubits 0 and 1. It can either act as the identity or as a $Z$ gate on qubit 0 via phase kickback. The second is a CNOT between qubits 2 and 0, which acts either as the identity or an X-gate on qubit 0. Whether these two gates act as the identity or a Pauli gate depends on the specific software state used.\\
Consequently, the effect on qubit 0 is a superposition of the identity (no CNOTs activate), $Z$ (first CNOT activates), $X$ (second CNOT activates) and $Y$ (both CNOTs activate). This demonstrates that the circuit induces Pauli noise on the channel. A similar analysis shows that qubit 1 is also subject to Pauli noise with potentially different error rates.

\subsection{Optimal Niu--Griffiths programs are real-valued}
The most general software state $\ket{\psi}$ to program the NG cloner is given by
\begin{equation} \label{software}
\ket{\psi} =
\begin{pmatrix}
a e^{i \alpha}\\ b e^{i \beta}\\ c e^{i \gamma}\\ d e^{i \delta}
\end{pmatrix}.
\end{equation}
Note that the circuit in Figure~\ref{1qubitNGcircuit}  only generates real states with $\alpha=\beta=\gamma=\delta=0$. We will justify this choice later in this section.\\
We can treat the circuit as a unitary transformation and calculate fidelities of the two output clones (qubits 0 and 1) relative to the states prepared by Alice. For this analysis, we consider the basis states of the $Z$, $X$ and $Y$ bases,
\[
\{\ket{0}, \ket{1}\}, \quad \{\ket{+}, \ket{-}\}, \quad \{\ket{+i}, \ket{-i}\}.
\]
The fidelity can then be computed for each state individually. For example, for state $\ket{0}$, the quantum fidelity with Bob's density matrix $\rho_B$ is given by
\begin{equation}
F_{AB, \ket{0}} = \bra{0} \rho_B \ket{0}.
\end{equation}
Inspection reveals that the NG cloner clones each state within the same basis with the same fidelity, allowing to define a single fidelity for each basis. For example for basis $Z$, $F_{AB, Z} = F_{AB, \ket{0}} = F_{AB, \ket{1}}$.
The fidelities can then be expressed directly as functions of the software state coefficients $a$, $b$, $c$ and $d$ and the angles $\alpha, \beta, \gamma, \delta$ as follows:
\begin{eqnarray}\label{NG1qubitfidelities}
F_{AB,Z} &=& a^2 + c^2 \label{NGfbz}\\
F_{AB,X} &=& a^2 + b^2 \label{NGfbx}\\
F_{AB,Y} &=& a^2 + d^2 \label{NGfby}\\
F_{AE,Z} &=& 1/2 + ac \cos(\alpha - \gamma) + bd \cos(\beta - \delta)\label{NGfez}\\
F_{AE,X} &=& 1/2 + ab \cos(\alpha - \beta) + cd \cos(\gamma - \delta)\label{NGfex}\\
F_{AE,Y} &=& 1/2 + ad \cos(\alpha - \delta) + bc \cos(\beta - \gamma)\label{NGfey} 
\end{eqnarray}

Additionally, normalization of the software state gives:
\begin{equation}\label{norm}
a^2 + b^2 +c^2 +d^2 = 1.
\end{equation}
We define an \textit{optimal} software as one that maximizes a weighted sum of fidelities. We define the quality of the cloner for Bob $$ Q_B = x F_{ABX} + y F_{ABY} + z F_{ABZ}$$ for $x,y,z$ in the interval $[0,1]$, and similarly $Q_E$ for Eve. For a fixed $Q_B$, the optimization task is to find $a,b,c,d$ and $\alpha,\beta,\gamma,\delta$ that maximize $Q_E$. For example, $x=y=z$ yields the UQCM, while for $x=y, z=0$  we find the PCCM in the $XY$ plane.\\
To justify that we only consider real program states, we observe that a maximization of Eve's fidelity in one of the bases is achieved if two of the angles $\alpha$, $\beta$, $\gamma$, $\delta$ are equal. As soon as one maximizes at least two of Eve's fidelities simultaneously, all four angles have to be equal, and we can always choose them to be zero. Therefore, it is sufficient to consider only real coefficients, and the quantum circuit in Figure~\ref{1qubitNGcircuit} can implement all optimal NG cloners.\\
Note that our implementation in Figure~\ref{1qubitNGcircuit} differs from the original circuit in~\cite{niugriffiths} by including a conveniently parameterized state preparation. With the three real parameters $\rho$, $\phi$ and $\theta$, the circuit can implement all real-valued program states. We can think of one parameter as responsible for controlling the asymmetry between the fidelities of Bob's and Eve's clones. The remaining two parameters can be used to characterize a family of NG cloners with bias between the fidelities in the three different bases.\\
This analysis applies equally to the single-qubit QID, for which we replicate the results in Appendix~\ref{QIDsinglequbit}. While the QID and NG cloners possess equivalent capabilities for a single qubit, their performance differ for multi-qubit states. Specifically, higher-dimensional QIDs act as Heisenberg cloners~\cite{cerfAsymmetric}, whereas we focus on Pauli cloners.

\subsection{Optimal Niu--Griffiths cloning for single qubits with Pauli noise}
The Niu--Griffiths cloner can implement a broad range of useful cloners. As noted in~\cite{qkdasQML}, when errors affect the two bases of the BB84 protocol differently, so-called \textit{imbalanced cloners} can yield better average fidelities than the PCCM, which is the optimal attack on the protocol in the noiseless case. We show here that the NG cloner can replicate these imbalanced cloners and extend this analysis to the three bases of the six-state protocol. Adopting an application-oriented perspective, we seek to identify the optimal cloners when the qubits are subject to noise. To this end, we consider a general Pauli noise model defined by:
\begin{equation}
\begin{split}
\mathcal{E}(\rho) ={}& (1 - p_X - p_Y - p_Z) \, \rho \\
& + p_X \, X \rho X^\dagger \\
& + p_Y \, Y \rho Y^\dagger \\
& + p_Z \, Z \rho Z^\dagger.
\end{split}
\end{equation}

The impact of such a model on the fidelities in the $X$, $Y$ and $Z$ bases is derived in Appendix~\ref{appendix:noise_effect}. The resulting fidelities $\tilde{F}$ are given by:
\begin{align}
\tilde{F}_{AB,X} &= F_{AB,X}(1 - 2p_Y - 2p_Z) + p_Y + p_Z,\label{eq:NGnoisy_fabx} \\
\tilde{F}_{AB,Y} &= F_{AB,Y}(1 - 2p_X - 2p_Z) + p_X + p_Z,\label{eq:NGnoisy_faby} \\
\tilde{F}_{AB,Z} &= F_{AB,Z}(1 - 2p_X - 2p_Y) + p_X + p_Y,\label{eq:NGnoisy_fabz}
\end{align}
and similarly for the fidelities between Alice and Eve.\\

\subsubsection{BB84 imbalanced cloner}
\begin{figure}[!h]
\begin{center}
\includegraphics[width=0.9\columnwidth]{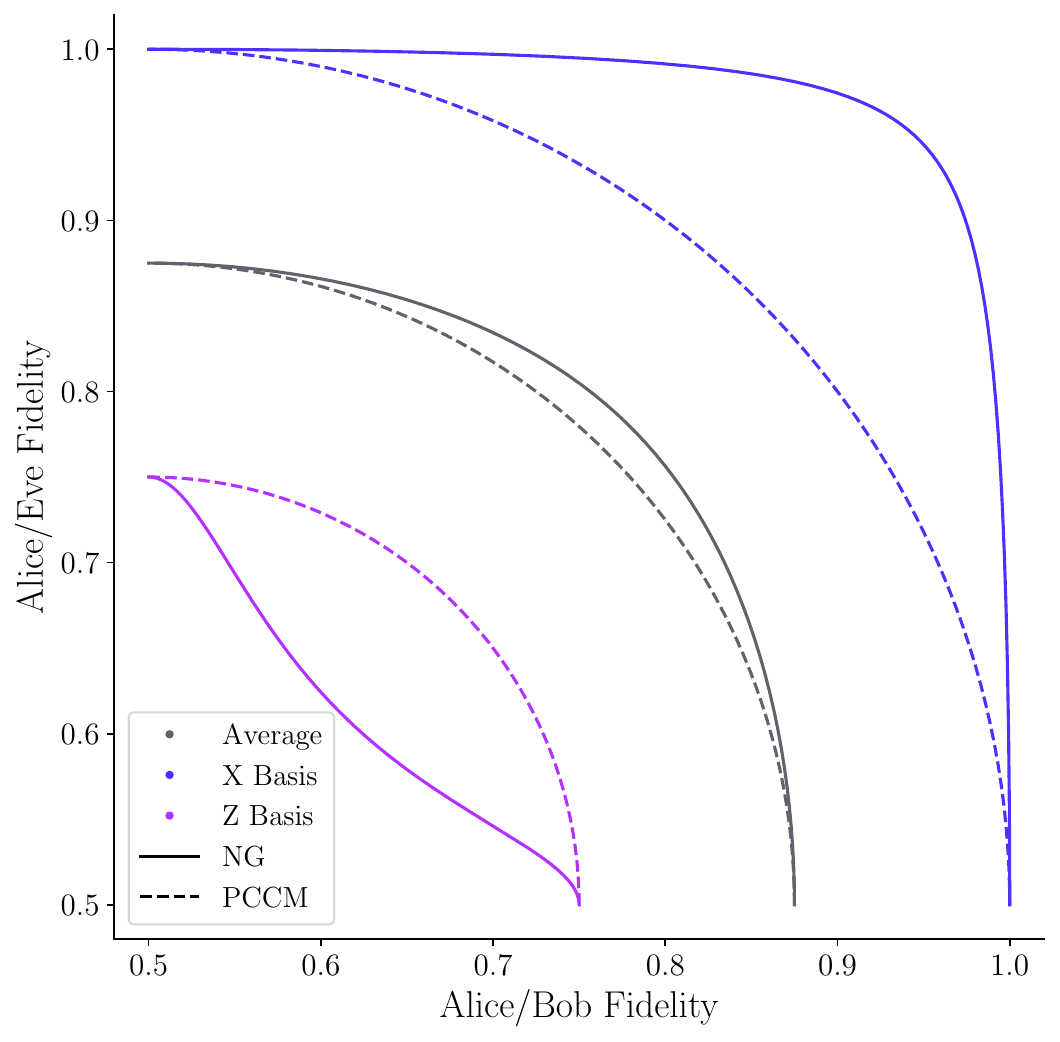}
\end{center}
\caption{Comparison between the PCCM and the imbalanced cloner obtained with Niu--Griffiths for a noisy channel with $p_X = 0.25$. The average fidelity is displayed alongside the specific fidelities for the $X$ and $Z$ bases. By biasing the cloner towards the $X$ basis which remains invariant under the bit-flip noise, the Niu--Griffiths cloner achieves a higher average fidelity than the PCCM.}
\label{fig:zxnoise}
\end{figure}

Using the optimization framework previously described, we can identify the optimal program states for a given noise model. For instance, we consider the same example as in~\cite{qkdasQML} where the BB84 protocol is subject to bit-flip noise ($X$) with probability $p_X = 0.25$. The best fidelities obtained with the NG cloner are displayed in Figure~\ref{fig:zxnoise} with the PCCM as reference. By strategically reducing fidelity in the noise-affected $Z$ basis, the NG cloner gains more in the unaffected $X$ basis, thereby outperforming the PCCM on average.

\subsubsection{Generalized imbalanced cloners}

\begin{figure}[!h]
\begin{center}
\includegraphics[width=0.9\columnwidth]{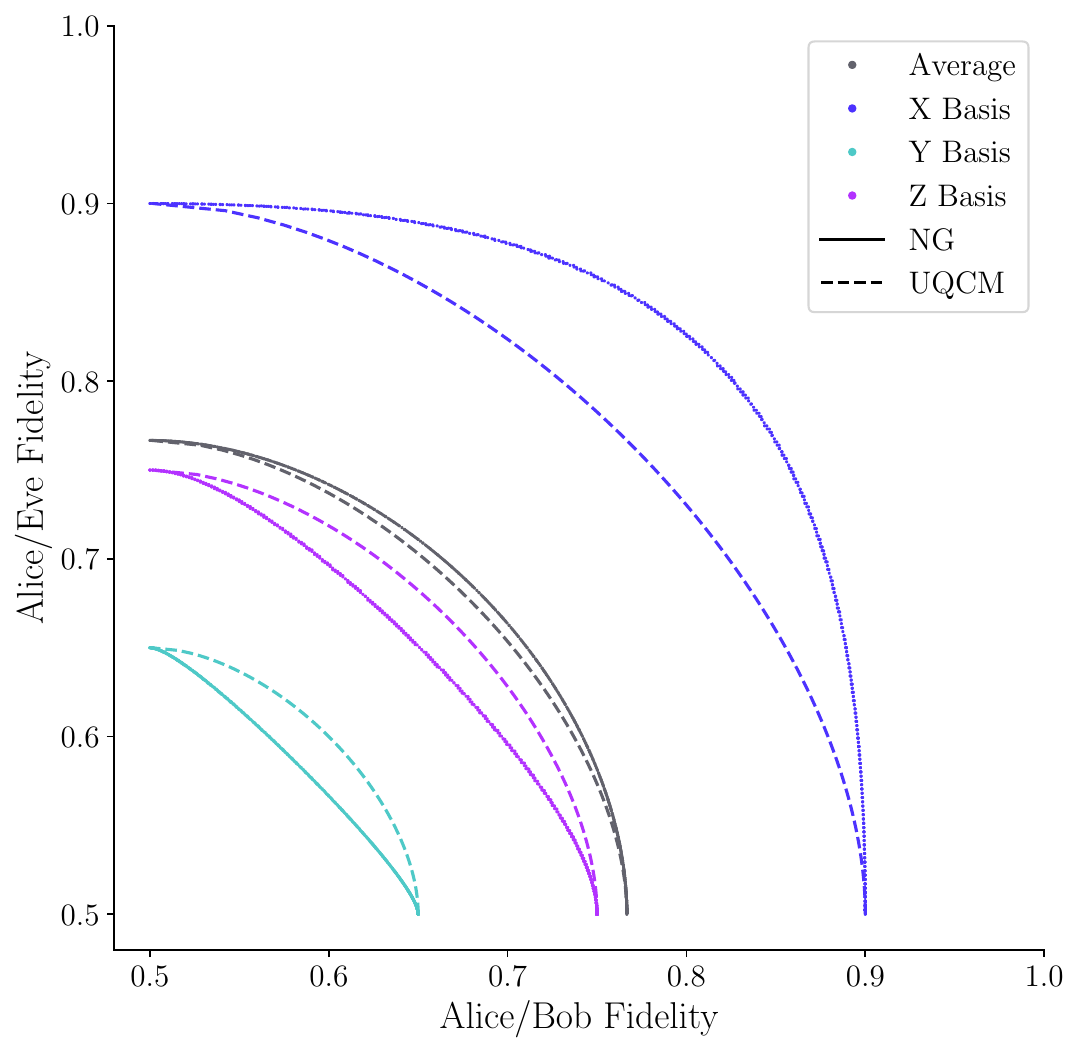}
\end{center}
\caption{Comparison between the UQCM and the generalized imbalanced cloner obtained with Niu--Griffiths for a noisy channel with $p_X = 0.25$ and $p_Z = 0.1$. The average fidelity is displayed alongside the specific fidelities for the $X$, $Y$ and $Z$ bases. By biasing the cloner towards the $X$ basis which is only affected by the weaker $Z$ errors, the Niu--Griffiths cloner achieves a higher average fidelity than the UQCM.}
\label{fig:NGsixstatenoise}
\end{figure}

We now extend this analysis to the three bases of the six-state protocol. As this set is maximal in terms of mutually unbiased bases (MUBs), it provides the most general scenario for single-qubit cloning under our noise model.
In this broader setting, the additional complexity introduced by the third basis complicates an analytical analysis such as in~\cite{qkdasQML}. Instead, we rely on numerical methods to explore the achievable fidelities. Our results demonstrate that the known optimal attack for the six-state protocol, the UQCM, can be outperformed for specific values of $(p_X, p_Y, p_Z)$. More generally, the numerical evidence suggests that whenever a noise model introduces an imbalance between the three bases $X$, $Y$ and $Z$, it is possible to outperform the UQCM. An example of such a noise model is displayed in Figure~\ref{fig:NGsixstatenoise}. We expect these biased cloners to be optimal in the sense that higher average fidelities cannot be achieved by any unitary for the noise models they are associated with. However, a formal proof of optimality is not provided here. A brief discussion on how this proof could be achieved is included in the conclusion of this paper.

The fact that the UQCM can be outperformed for many choices of Pauli noise is in fact not particularly surprising. Consider, for instance a noise model with $p_Z = p_X = 0.25$ and $p_Y = 0$. In that case, the fidelities of Bob and Eve in the $Y$-basis are fixed at $F_{AB,Y} = F_{AE,Y} = 1/2$ regardless of the software state $\ket{\psi}$, while the fidelities in the $Z$ and $X$ bases are not fixed but affected equally. This situation is therefore equivalent to a noisy BB84 protocol without imbalance between the two bases. In this setting, it is well known that the PCCM is the optimal cloner and outperforms the UQCM. If we keep $p_Y = 0$ but now allow $p_Z \neq p_X$ with $p_Z + p_X = 0.5$, the problem reduces to a noisy version of BB84 where the corresponding imbalanced cloner outperforms the PCCM. The PCCM and imbalanced cloners are all special cases of these more general biased cloners.

\subsection{Examples of optimal one qubit NG programs}
\label{sec:NG1qb}
It is convenient to parameterize the NG software state $\ket{\psi}$ in terms of the three rotation angles used in Figure~\ref{1qubitNGcircuit}. In the computational basis, the state is represented by the column vector:
\begin{equation} \label{NGprogram}
\ket{\psi} =
\begin{pmatrix}
a \\ b \\ c \\ d
\end{pmatrix} =
\left(
\begin{array}{c}
\cos{\theta} \cos{\rho} \\[0.5mm]
\cos{\phi} \sin{\rho} \\[0.5mm]
\sin{\theta} \cos{\rho} \\[0.5mm]
\sin{\phi} \sin{\rho}
\end{array}
\right) .
\end{equation}
By optimizing for specific cloning problems as outlined above, we derive explicit parametrizations for several types of cloners, presented in Table~\ref{NGcloners}. For the UQCM and PCCM, one free parameter controls the asymmetry between the two copies. The imbalanced cloner described in~\cite{qkdasQML} includes an additional parameter to account for basis biases.
\begin{table}[h]
    \centering
    \begin{tabular}{c|c|c}
       Cloner & Angles & sym.\\
       \hline
       UQCM & $\phi = \pi/4, \rho = \arctan(\sqrt{2} \sin\theta )$ & $\theta= \arctan(1/3)$\\
       PCCM & $\phi = \rho = \theta$ & $\theta=\pi/8$\\
       Imb. Cloner & $\phi = \theta , \rho = \arctan(\eta \tan2\theta)/2$ & $\theta = \pi/8$
    \end{tabular}
    \caption{Parametrization of some standard cloning machines using the Niu--Griffiths cloner. The angles $\phi$, $\rho$ and $\theta$ are those of the circuit in Figure~\ref{1qubitNGcircuit}.
    The parameter $\eta$ quantifies the noise imbalance within the channel. Under our noise model, it is defined as $\eta = (1 - 2p_Z - 2p_Y)/(1 - 2p_X - 2p_Y)$. Note that when $p_Z = p_X$, it simplifies to $\eta = 1$ and we find the same parametrization as the PCCM.
    The PCCM and the imbalanced cloner are for the bases of the BB84 protocol, i.e. $X$ and $Z$. One can also implement parameterizations for the $XY$ and $YZ$ bases. The column \textit{sym} gives the value for the symmetric cloner.}
    \label{NGcloners}
\end{table}

\subsection{Cloning machines that cannot be implemented with the Niu--Griffiths cloner or QID}
\begin{figure}[!h]
\begin{center}
\includegraphics[width=0.9\columnwidth]{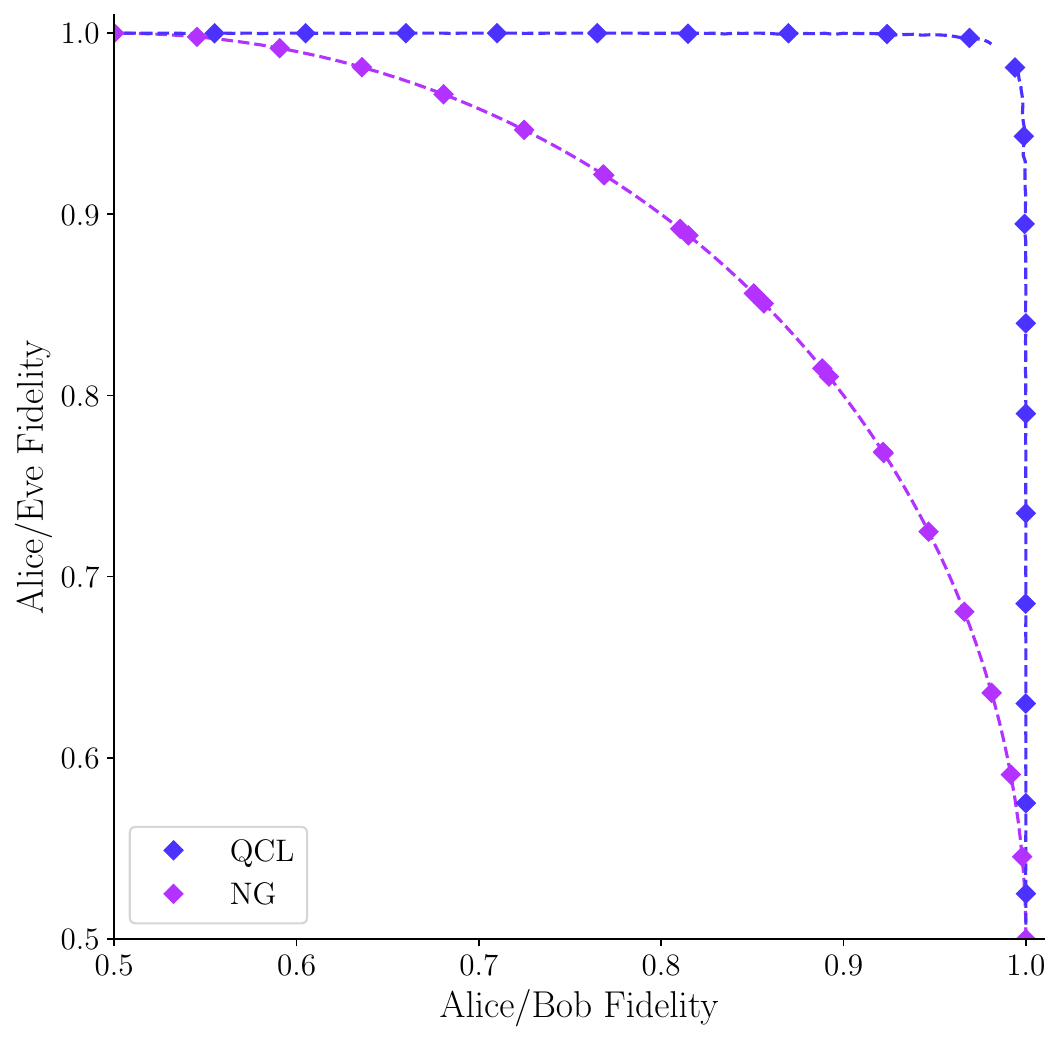}
\end{center}
\caption{Comparison between the best Niu--Griffiths cloner and QML results for cloning the two states of the B92 protocol. QML achieves significantly higher fidelities than the NG cloner. The QID performance (not shown here) is equal to that of the NG cloner).}
\label{b92}
\end{figure}

From the results of the previous section, it is tempting to assume that the Niu--Griffiths cloner can implement any cloning machine, or maybe any optimal cloning machine. However, we demonstrate here that this is not the case.\\
As an example, we consider the task of cloning only the states $\ket{0}$ and $\ket{+}$ used in the B92 protocol~\cite{B92}. Unlike BB84 or six-state protocols, which utilize full bases, B92 relies only on two non-orthogonal states, each chosen from two different MUBs.\\
We use the QML approach detailed in~\cite{qkdasQML} to optimize the cloning of these two states. We use a parametrized circuit and optimize its parameters with gradient-based methods to maximize the fidelities of Alice and Bob. For more details on the ansatz, cost function and optimizer, see Appendix~\ref{appendix:qcl}. The optimization results are shown in blue in Figure~\ref{b92}. To identify the best cloners achievable with the NG cloner (and also with the QID, which leads to the same results), we perform a grid search over the amplitudes of the software state. The best results, shown in purple, are compared with those obtained by QML, and appear to be significantly lower.

A theoretical treatment of this cloning problem is provided in~\cite{stateDep}, where it is shown that such cloning machines copy the two states in each basis with different fidelities. In contrast, the 1-qubit NG cloner (and also the 1-qubit QID) can only clone the two states of each basis with identical fidelity. Consequently, the best a NG cloner or QID can achieve is to maximize the fidelities in the two bases from which the B92 states are picked. These bases are exactly those of the BB84 protocol, for which the optimal attack is known to be the PCCM. Therefore, the NG cloner and the QID cannot clone the B92 states optimally.

\section{Cloning Pauli channels with $N \geq 2$}
Commercial QKD systems utilizing single-qubit channels are already available, making the single-qubit cloners discussed before of immediate practical interest. First steps towards real world application were taken in~\cite{qinutshell}, where imbalanced cloners were implemented on an ion-trap-based quantum computer.\\
For improved security and higher data rates, a natural step is to move from single qubits to larger registers. This applies to QKD, quantum communication, and even applications such as quantum money~\cite{wiesner1983}. In these higher-dimensional setups, cloners remain central to analyzing potential attacks.\\
As long as the quantum states are free of noise, existing cloners, such as the UQCM or the PCCM, will be practical as well as optimal in many of such new setups, for example for the construction of attacks. Simple circuit constructions are available, for example the UQCM for arbitrary qudits with the QID~\cite{QID}. In the presence of noise, however, the situation is different. As Cerf pointed out in~\cite{cerfAsymmetric}, higher dimensional QIDs are Heisenberg cloners. We will see that they are not optimal cloners for general Pauli channels.
We provide a detailed analysis in Appendix~\ref{appendix:20fids}, where we show that the QID cannot introduce biases between a certain pair of MUBs.
Furthermore, not all states in each basis are cloned with equal fidelity.
\\
Consequently, the construction of higher dimensional Pauli cloners seems of interest.
\subsection{MUBs for two-qubit states} \label{2qubitMUBs}
As a concrete example, we first consider the case of two-qubit channels. The security of the BB84 and six-state protocols, as well as the idea of quantum money~\cite{wiesner1983}, relies on the fundamental property of mutually unbiased bases: a state prepared in one basis will, when measured in any other basis, yield all possible outcomes with equal probability. For a brief review of MUBs, see e.g.~\cite{MUBs3ways}.\\
These principles extend naturally to $N$-qubit registers. It is well known that there are $2^N+1$ MUBs for $N$-qubit states. For two-qubit channels ($N=2$), these 5 MUBs are unique~\cite{MUBs}. Since each basis contains four states, these five bases can be used to define a 20-state QKD protocol, which would be the two-qubit equivalent of the six-state protocol. \\
To facilitate numerical simulations, we explicitly construct the bases $M_0,M_1,M_2,M_3,M_4$ as follows:

{\small 
\allowdisplaybreaks
\begin{align}
M_0 &=
\vector 1\\0\\0\\0.,
\vector 0\\1\\0\\0.,
\vector 0\\0\\1\\0.,
\vector 0\\0\\0\\1.,\\
M_1 &= 
\frac{1}{2}\vector 1\\1\\1\\1.,
\frac{1}{2}\vector 1\\-1\\1\\-1.,
\frac{1}{2}\vector 1\\1\\-1\\-1.,
\frac{1}{2}\vector 1\\-1\\-1\\1.,\\
M_2 &= 
\frac{1}{2}\vector 1\\-1\\i\\i.,
\frac{1}{2}\vector 1\\1\\i\\-i.,
\frac{1}{2}\vector 1\\1\\-i\\i.,
\frac{1}{2}\vector 1\\-1\\-i\\-i.\\
M_3 &= 
\frac{1}{2}\vector 1\\ i\\ i\\ -1.,
\frac{1}{2}\vector 1\\ -i\\ i \\ 1.,
\frac{1}{2}\vector 1\\ i\\ -i\\ 1.,
\frac{1}{2}\vector 1\\ -i\\ -i\\ -1.,\\
M_4 &=
\frac{1}{2}\vector 1\\ -i\\ -1\\ -i., 
\frac{1}{2}\vector 1\\ i\\ 1\\ -i.,
\frac{1}{2}\vector 1\\ -i\\ 1 \\ i.,
\frac{1}{2}\vector 1\\ i\\ -1\\ i..
\end{align}
}
With quantum circuits, these states are prepared by initializing the two qubits in states $\ket{00}$, $\ket{01}$, $\ket{10}$ and $\ket{11}$ and then applying the following unitary operations: 

\begin{description}
    \item[$M_0$] $I \otimes I$
    \item[$M_1$] $H \otimes H$
    \item[$M_2$] $(H \otimes H) \to (S \otimes S) \to \text{CNOT}_{1 \to 0}$
    \item[$M_3$] $(H \otimes H) \to (S \otimes S)$
    \item[$M_4$] $(H \otimes H) \to (S \otimes S) \to (Z \otimes Z) \to \text{CNOT}_{0 \to 1}$
\end{description}

\subsection{Pauli channels and cloners for $N$ qubits} \label{paulichannels}
We now assume Alice communicates with Bob via an N-qubit Pauli channel. The channel acts on the N-qubit state $\rho$ by:
\begin{equation} \mathcal{E}(\rho) = \sum_{i \in \mathcal{P}_N} p_i P_ i\rho P_i,
\end{equation}
where $P_i \in P_N = \{I, X, Y, Z\}^{\otimes N}$ are the Pauli operators and $p_i$ are the associated error rates with $\sum_i p_i = 1$. In particular, $p_{I^{\otimes N}}$ denotes the probability of error-free transmission. In this general channel, there are $4^N-1$ possible errors.

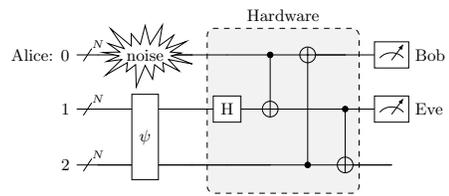
\begin{figure}[h!]
\centering
\begin{tikzpicture}
\node[scale=0.7]{
\tikzset{
noisy/.style={starburst,fill=white,draw=black,line
width=0.5pt,inner xsep=-4pt,inner ysep=-3pt}
}
\begin{quantikz}[thin lines, classical gap=0.09cm, column sep = 0.4cm, color=black,background color=white]
\lstick{Alice: 0}&\qwbundle{N} &\gate[1,style={noisy},label style=black]{\text{noise}} &\qw  &  \gategroup[3,steps=4,style={dashed,rounded corners,fill=black!5, inner xsep=0pt},background]{Hardware} & \ctrl{1} & \targ{} &\qw &\meter{} \rstick{Bob} \\
\lstick{1}&\qwbundle{N} & \gate[2]{\psi} &  \qw & \gate{\text{H}} & \targ{} & \qw & \ctrl{1} &\meter{} \rstick{Eve} \\
\lstick{2}&\qwbundle{N} & \qw &  \qw & \qw & \qw & \ctrl{-2}&\targ{} &\qw
\end{quantikz}
};
\end{tikzpicture}
    \caption{Quantum circuit implementing the Niu--Griffiths cloner for $N$-qubit registers. Register 0 serves as both input and first output, while register 1 is the second output. Register 0 can be affected by Pauli noise, but only before Eve's controlled gates begin to act. The $H$ gate symbol stands for $N$ single-qubit $H$ gates, such that one H gate acts on each qubit in the register. All symbols for CNOT gates in the diagram  correspond to $N$ CNOT gates, with the $i$-th CNOT having the control on the $i$-th qubit of the control register, and the target in the $i$-th qubit of the target register.}
    \label{NqubitNGcircuit}
\end{figure}

We now generalize the Niu--Griffiths construction to $N$-qubits, such that they keep the defining property that they introduce Pauli errors in the quantum channel used by Alice and Bob. We will then see that this has a number of desirable properties. For instance, we will show that it offers the freedom to introduce biases between each MUB while cloning each state within a basis with the same fidelity. This enables to produce analogues of the imbalanced cloners described in~\cite{qkdasQML}.\\
The construction is straightforward: each qubit of the circuit in Figure~\ref{1qubitNGcircuit} is replaced by a register of $N$ qubits, totaling $3N$ qubits.
The state preparation can also be generalized to $N$ qubits with multi-controlled rotations, but the complexity of this construction grows exponentially with $N$. Consequently, for larger systems, one may want to restrict the state preparation to the most relevant cases.
\\One crucial question is the generalization of CNOT gates. They are applied bit-wise between the two registers, such that the $i$-th qubit of the controlling register acts on the $i$-th qubit of the target register. This configuration, shown in Figure~\ref{NqubitNGcircuit} ensures that the circuit can now generate all Pauli operations on each bit of cloned register, just as in the single-qubit case.

\begin{figure}[!h]
\begin{center}
\begin{tikzpicture}
\node[scale=0.7]{
\begin{quantikz}[thin lines, classical gap=0.2cm, column sep = 0.4cm]
\lstick{Alice: 0} & \qw & \qw & \qw \gategroup[4,steps=2,style={dashed,rounded corners,fill=blue!10, inner xsep=0pt},background]{} & \ctrl{2} & \qw \gategroup[6,steps=2,style={dashed,rounded corners,fill=blue!10, inner xsep=0pt},background]{}& \targ{}  & \qw \gategroup[6,steps=2,style={dashed,rounded corners,fill=blue!10, inner xsep=0pt},background]{} & \qw & \rstick[2]{Bob}\\
\lstick{Alice: 1} & \qw & \qw & \ctrl{2} & \qw & \targ{} & \qw & \qw & \qw & \qw \\
\lstick{2} & \gate[4]{\makebox[1.5em]{$\ket{\psi}$}} & \gate{\text{H}} & \qw & \targ{}  \qw & \qw & \qw & \qw & \ctrl{2} & \rstick[4]{Eve}\\
\lstick{3} & \qw & \gate{\text{H}} & \targ{}  & \qw & \qw & \qw & \ctrl{2} & \qw & \\
\lstick{4} & \qw & \qw & \qw & \qw & \qw & \ctrl{-4}  & \qw & \targ{} & \qw\\
\lstick{5} & \qw & \qw & \qw & \qw & \ctrl{-4}  & \qw & \targ{} & \qw & \qw
\end{quantikz}
};
\end{tikzpicture}
\end{center}
\caption{Circuit implementation of the Niu--Griffiths cloner for two-qubit registers. The state $\ket{\psi}$ on qubits 2 to 5 programs the circuit for a given task. The hardware part is given by bitwise CNOTs, as opposed to the shift operators in the QID.}
\label{fig:2qubitNGcircuit}
\end{figure}
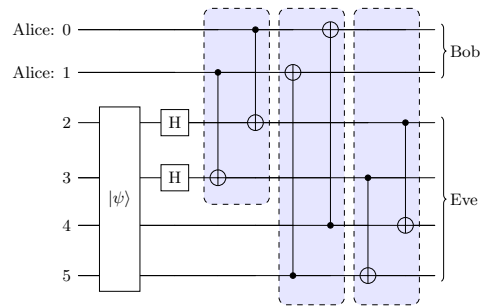

For clarity, we show the two-qubit version (which we will analyse further in Section~\ref{twoqubitcloners}) explicitly in Figure~\ref{fig:2qubitNGcircuit}.

\subsection{Pauli errors and MUBs for two qubits}
 We now examine the effect of the Pauli errors on the MUBs we defined in Section~\ref{2qubitMUBs}. A summary is given in Table~\ref{tab:2qbnoiseeffect}. Note that the Pauli errors form commuting sets (Abelian groups when the case of no error occurring, $I \otimes I$, is included) of three types of errors, where each of those groups leave one of the bases invariant (up to phases).

\begin{table}[!h]
    \centering
    \begin{tabular}{l|c|c|c|c|c}
       & $M_0$  &  $M_1$ & $M_2$ & $M_3$ & $M_4$\\
       \hline
       $X \otimes X$, $I\otimes X$, $X\otimes I$ & 1 & 0 & 1 & 1 & 1\\
       $X\otimes Z$, $Y\otimes X$, $Z\otimes Y$ & 1 & 1 & 1 & 1 & 0\\
       $Z\otimes X$, $X\otimes Y$, $Y\otimes Z$ & 1 & 1 & 0 & 1 & 1\\
       $Z\otimes Z$, $I\otimes Z$, $Z\otimes I$ & 0 & 1 & 1 & 1 & 1\\
       $Y\otimes Y$, $I\otimes Y$, $Y\otimes I$ & 1 & 1 & 1 & 0 & 1
    \end{tabular}
    \caption{2-qubits Pauli errors and their effect on each of the five MUBs. A \textit{1} means that the error permutes the basis states while \textit{0} indicates that each state stays invariant (up to a phase). Each MUB is immune to three kinds of pairs of Pauli errors. Each of these triplets of errors, together with the identity, forms an Abelian group.}
    \label{tab:2qbnoiseeffect}
\end{table}
To understand the action of the Pauli cloner, we examine the effect of Pauli error operators on the MUB states. For example, considering the three bit-flip errors in the first row of Table~\ref{tab:2qbnoiseeffect}, one can observe their action on the basis states of $M_0$. Specifically, each error induces a pairwise swap of the four basis vectors within $M_0$. There are three different ways to swap the four basis elements pairwise. We can denote these by (12)(34), (13)(24) and (14)(23). One finds that (12)(34) is implemented by $X \otimes I$, and (13)(24) by $I \otimes X$ and (14)(23) by $X \otimes X$. When applied to the states in bases $M_2, M_3, M_4$, one also finds that each error swaps the basis elements pairwise. In contrast, the states in $M_1$ remain unchanged, up to overall phases. One can check that the effects of all the other errors in the table also implement such pairwise swaps. The errors in each line of the table, together with the identity (i.e., no error occurring), form a Klein four group $Z_2 \times Z_2$~\cite{klein}. The structure summarized in the table implies an important connection between the cloning of qubits affected by Pauli noise, MUBs and biased cloners: each triplet of Pauli errors found in each line of Table~\ref{tab:2qbnoiseeffect} leaves the states of one MUB unaffected (up to a phase). Therefore, a cloner can trade extra fidelity in the unaffected basis for fidelity in other, affected bases, such that the average fidelity rises, compared to an unbiased cloner. This is exactly the effect described in~\cite{qkdasQML} for single qubits. The structures we described here for two qubits generalize to higher numbers of qubits.

\subsection{The role of all $4^N-1$ Pauli cloners}
In Section~\ref{1qubitNG}, we argued that in the single-qubit case, the three real parameters of Niu-Griffith cloners correspond to clone asymmetry and MUBs biases. Applying this logic to $N$ qubits, we find that there are many more free parameters ($4^N-1$ since the software register is $2N$ qubits) than MUBs biases ($2^N$). However, the number of parameters matches exactly the number of Pauli errors for $N$ qubits, and this is not a coincidence: we can think of the software state $\ket{\psi}$ as the instruction for which Pauli operators the cloner activates on each of the qubits to be cloned.\\
Note also, that the number of MUBs for $N$ qubits is $2^N+1$, and $\frac{4^N-1}{2^N+1} = 2^N-1$ is exactly the number of Pauli operators which leave the states in one MUB of the system unchanged (up to phases). These turn out to form a maximally commuting set (this was observed in~\cite{PauliMUBs} and used to construct MUBs) and since they (together with the identity, i.e. no error occurring) form an Abelian group in which every element squares to the identity, the group actually is isomorphic to $Z_2^{\otimes N}$. In the example $N=2$ we considered earlier, this means there are 5 copies of the Klein 4 group~\cite{klein}, one copy appearing in each line of Table~\ref{tab:2qbnoiseeffect}.\\
Finally, a comment on the use of the term "Pauli channel" in this work is in order: there are three different things which are Pauli channels: first, we allow the quantum channel which Alice and Bob use for communication to be affected by Pauli noise, so it is a Pauli channel. Second, even if the communication channel is noise free, the disturbance caused by a Pauli cloner causes Pauli noise in both clones it produces. So both outputs of the cloners are Pauli channels, too.\\
The power of Pauli cloners lies in the fact that they can be programmed to distribute the disturbance they introduce across all $4^N-1$ types of Pauli error that the communication channel can exhibit. In particular, we can tailor the noise they generate to the natural noise in the channel.
The most efficient way to do this in the case of no noise in the original channel (here, we measure this in terms of single copy fidelities, but this is true for other measures, like entropy, too) is to spread the disturbance evenly across all $4^N-1$ errors. This is what the UQCM does and is optimal in our sense as long as the Pauli error rates of the communication channel are all equal.\\
Another instructive example would be if the first bit in the communication channel would be affected by an $X$ error with 100\% probability: one can choose a program state for the cloner such that the disturbance is focused on an $X$ error on the first qubit, such that the two $X$ errors -- the one from noise in the original communication channel and the disturbance generated by the cloner -- cancel out. Such a cloner would actually improve the fidelity of the channel.\\
So Pauli cloners are defined by the $4^N$ dimensional state $\ket{\psi}$ which Eve prepares, and where each dimension corresponds to a Pauli error. The components of $\ket{\psi}$ control the disturbance of the corresponding Pauli error type that the cloner generates.

\begin{figure*}
\begin{center}
\includegraphics[width=1.6\columnwidth]{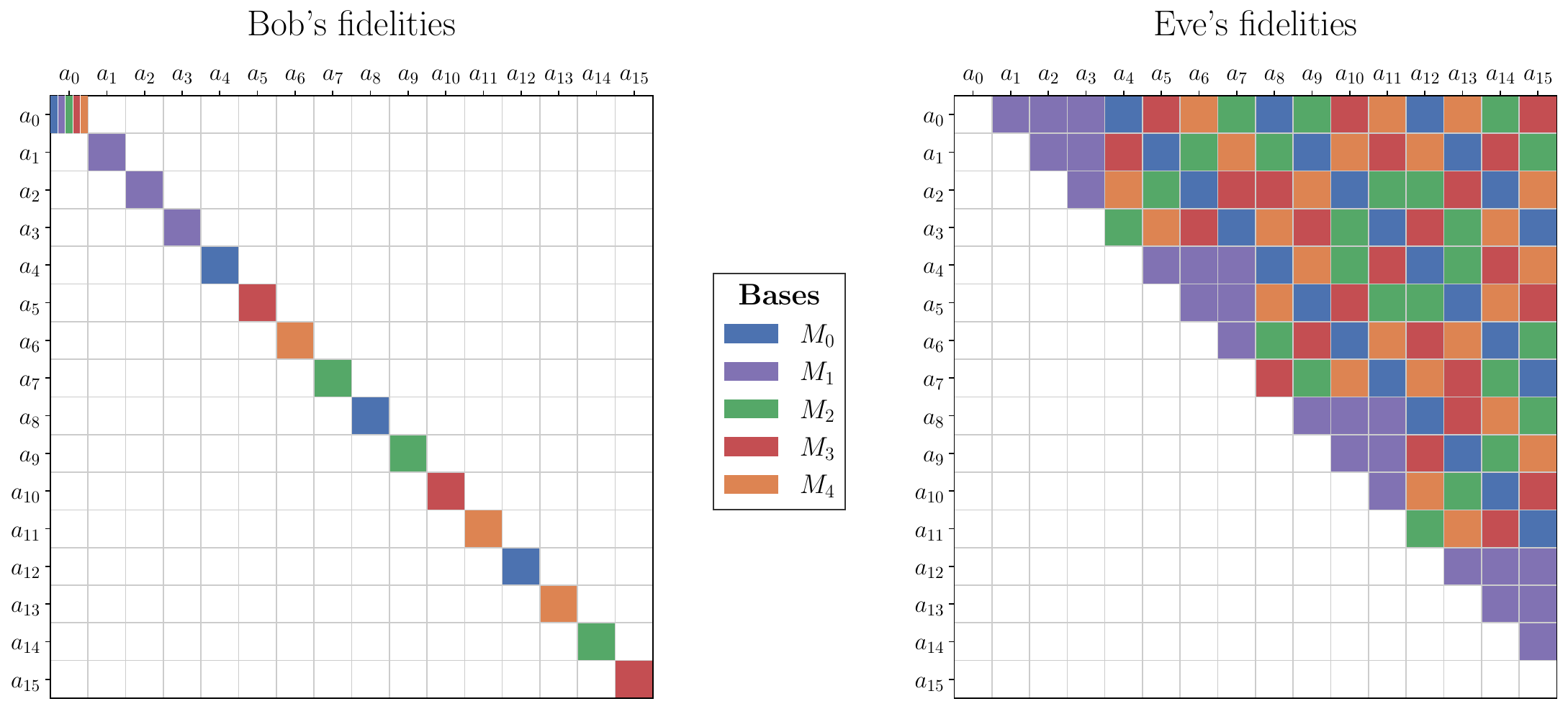}
\end{center}
\caption{Visual breakdown of the two-qubit fidelity terms. Bob's fidelities are shown in the left matrix, and Eve's in the right. Cell colors indicates for which basis the corresponding term contributes. Notably, $a_0^2$ (top left) is unique since it is the only term which appears more than once. It contributes to Bob's fidelity in all five bases.}
\label{fig:twoqubitfids}
\end{figure*}
\subsection{Analysis and examples of two-qubit Pauli cloner} \label{twoqubitcloners}

\subsubsection{Analytical fidelity expressions}
As in Section~\ref{1qubitNG}, we can treat the circuit in Figure~\ref{fig:2qubitNGcircuit} as a unitary and calculate the fidelities of the two output clones of Bob (qubits 0 and 1) and Eve (qubits 2 and 3) relative to the state prepared by Alice. The cloner is specified by the program state $\ket{\psi}$ initialized on qubit 2 to 5:
\begin{equation}
\ket{\psi}=\begin{pmatrix} a_0\\a_1\\ ...\\a_{15} \end{pmatrix},
\end{equation}
The resulting fidelity equations, detailed in Appendix~\ref{2qubitNGfidelities}, reveal the relationship between the Pauli channel induced by the cloner and the five MUBs. A key feature is that each basis fidelity is expressed as a sum of quadratic terms $a_ia_j$. Notably, each pair $(i, j)$ appears in exactly one fidelity expression except for $a_0^2$ which appears in each of Bob's fidelities. A visualization of the fidelities formulas is given in Figure~\ref{fig:twoqubitfids}. Each colored square represents one of the quadratic terms. For Bob's fidelity, only the diagonal terms $a_i^2$ appear.\\
As a special case, if $\ket{\psi} = \ket{0000}$ ($a_0 = 1$), Bob has a fidelity of 1 in each MUB while Eve's fidelity reduces to 0.5. Introducing a non-zero amplitude for $a_1$ means that qubit 5 is switched on, which, through the third CNOT gate, induces an $X$ error on qubit 1. Similarly, $a_2 \neq 0$ induces an $X$ error on qubit 0, while $a_3 \neq 0$ induces $X$ errors on both qubits 0 and 1.\\
Interestingly, Bob's fidelity for basis $M_1$, which is the only basis invariant under these three errors, is given by:
\begin{equation}
    F_{AB,M_1} = a_0^2 + a_1^2 + a_2^2 + a_3^2.
\end{equation}
The same analysis can be applied to the other four MUBs, which confirms that the three contributing coefficients for each basis correspond to binary encodings of the three errors associated with that MUB.

\subsubsection{Noisy two-qubit case}
\begin{figure}[!h]
\begin{center}
\includegraphics[width=0.9\columnwidth]{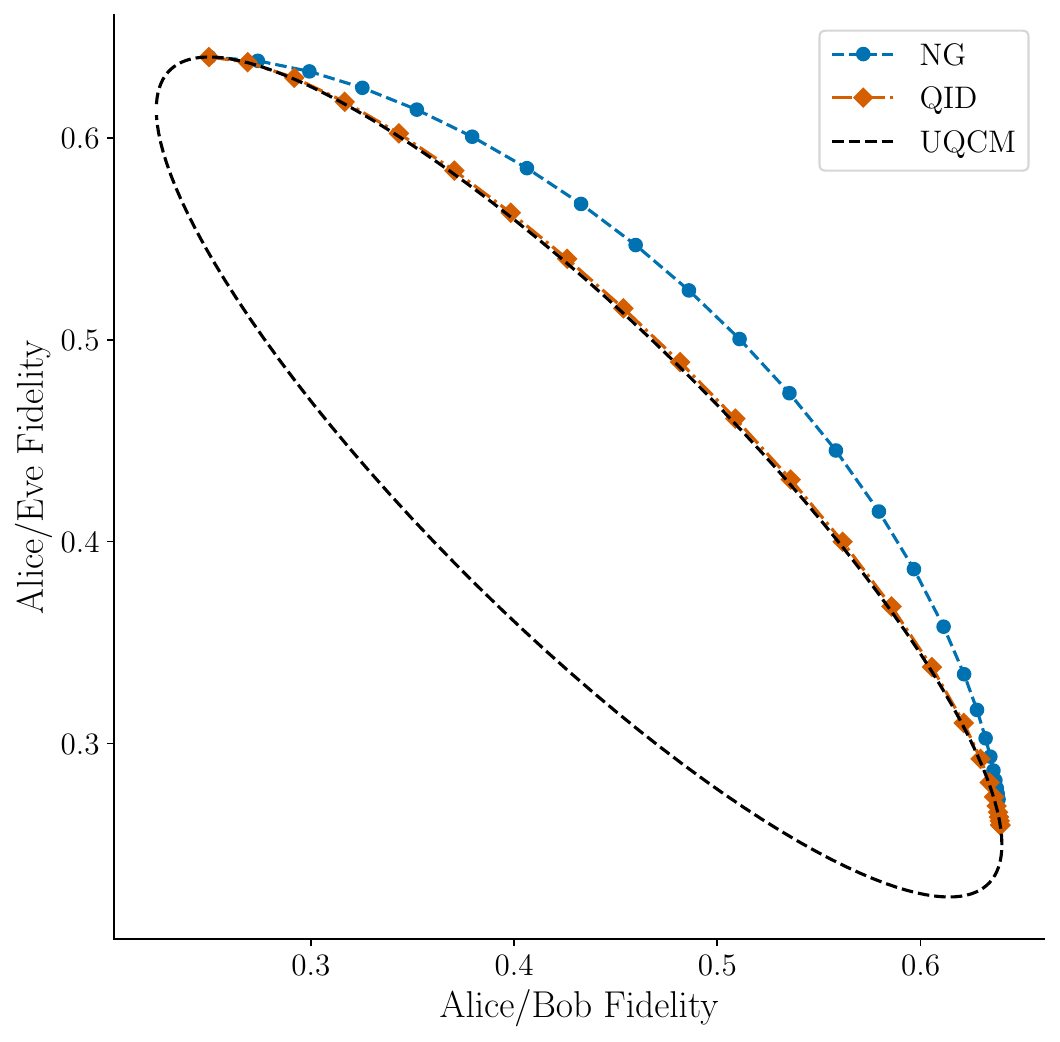}
\end{center}
\caption{Two-qubit cloning for maximal average fidelity over all MUBs when one of the two qubits is affected by Y-errors with an error rate of 45\%. The optimized NG cloner clearly outperforms the optimized QID. The performance of the UQCM is plotted as a reference.}
\label{ts_y_error}
\end{figure}

We now consider an example where the channel is affected by Pauli noise with a $45\%$ chance for a $Y\otimes I$ error. We use quantum machine learning on Eve's program state preparation (for details, see Appendix~\ref{twoqubitstateprep}) to maximize the fidelities of Alice and Bob.\\
As in the 1-qubit case, we find that the results clearly outperform the UQCM. By strategically biasing towards the unaffected $M_3$ basis, the cloner achieves a higher fidelity on average than the UQCM.\\
To further illustrate that the two-qubit NG cloner is naturally suited for Pauli noise, we also optimized the QID for the same problem. As illustrated in Figure~\ref{ts_y_error}, the QID is unable to match the fidelities of the NG cloner, as it is unable to adapt to any Pauli biases.

\begin{figure*}
\begin{center}
\includegraphics[width=1.6\columnwidth]{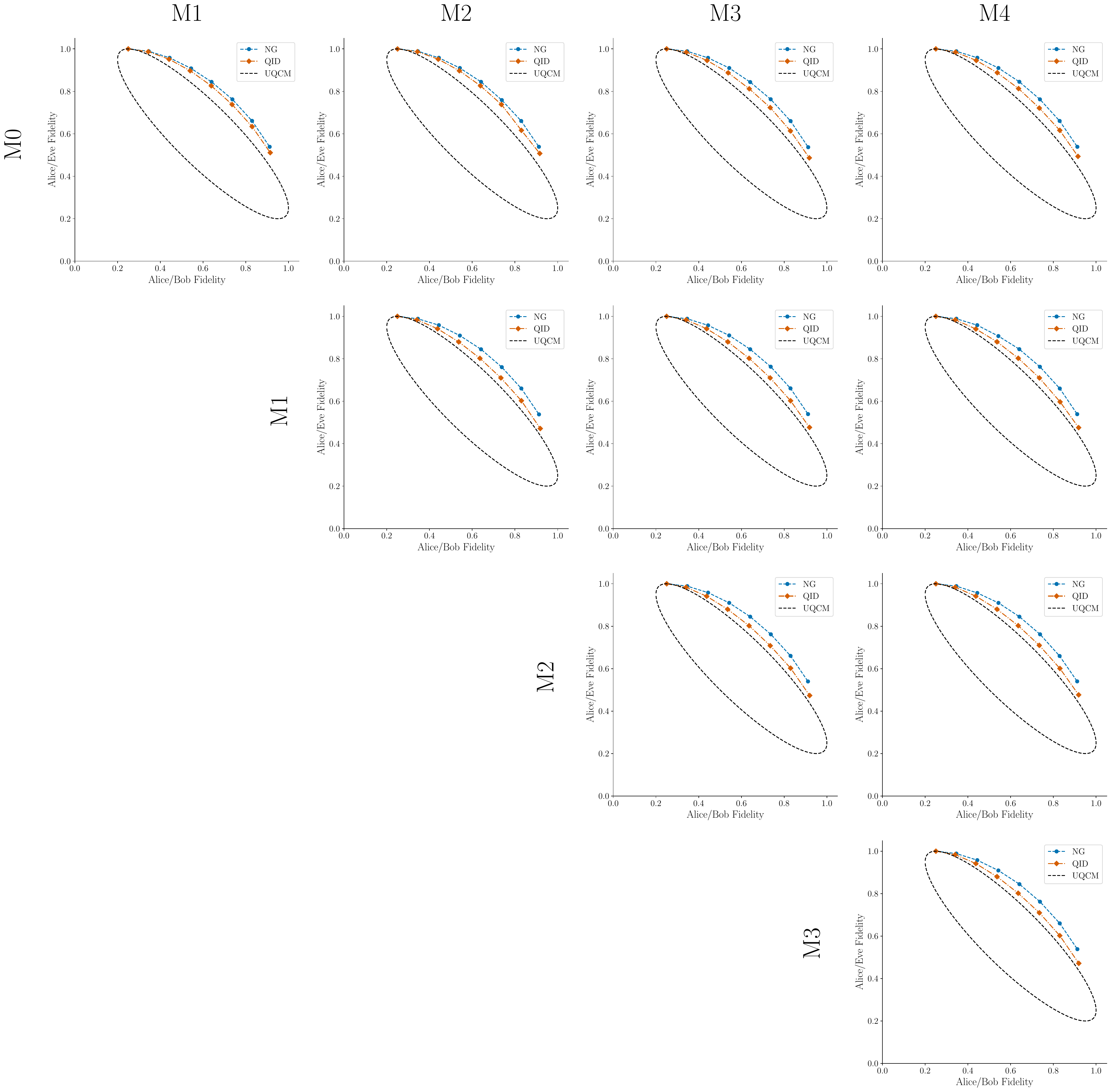}
\end{center}
\caption{Optimization of the two-qubit Niu--Griffiths cloner for reduced protocols utilizing two of the five available MUBs. Each subplot show the fidelities for a given pair of MUBs: $\{M_0, M_1\}$ (top-left),  $\{M_3, M_4\}$ (bottom-right), and so on. For comparison, the fidelities of the UQCM and optimized two-qubit QID are also plotted. These results illustrate that the performance of the Niu--Griffiths cloner is the same for any choice of basis pairs, while the QID performance depends on the choice of basis the pair.}
\label{sdf}
\end{figure*}

\subsubsection{Reduced two-qubit protocol}
Just as the BB84 protocol utilizes a subset (two of three) of the available single-qubit MUBs, one can define two-qubit protocols using only two of the five possible MUBs. We compared the optimization of the NG cloner and the QID across all ten possible pairs of bases. The resulting fidelities are presented in Figure~\ref{sdf}. The NG cloner consistently surpasses the QID in every scenario. Notably, the Pauli cloner performs identically, regardless of the choice of basis pair. In contrast, the QID performance varies with the choice of basis pair.

The two-qubit cloning examples we showed are, of course, far from exhaustive. One can optimize the Pauli cloner for any subset of bases, while at the same time we can switch on Pauli noise with any set of error rates for each of the 15 different Pauli errors. The machine learning techniques we describe in Appendix~\ref{twoqubitstateprep} are a convenient way to solve the cloning problem for any of those cases.

\subsection{Analytic version of the symmetric $N$-qubit UQCM implemented by a Pauli cloner}\label{NqubitUQCM}
The structure of the fidelity equations described previously (see Figure~\ref{fig:twoqubitfids}) generalizes straightforwardly to the general case of $N$ qubits: Bob's fidelity for basis $M_i$ is given by:
\begin{equation}
    F_{M_{i}} = a_0^2 + \sum_j a^2_j
\label{eq:generalized_f_b}
\end{equation}
where the sum is over $2^N-1$ terms from an index set which depends on $M_i$. Note that each terms $a_j^2$ $(j>0)$ contributes to only one $M_i$. So, for example, for $N=3$, we would have 9 MUBs, each associated with 7 different Pauli errors.\\
This basis-dependent architecture can also be specialized to implement the $N$-qubit symmetric UQCM which is characterized by a single fidelity $F$ which is the same across all bases and identical between the two clones. For $1\rightarrow 2$, the fidelity is given by~\cite{bucekhillery1998, cerf1998, werner1998}:
$$
F = \frac{d+3}{2(d+1)},
$$
with $d=2^N$. 
The structure of the Pauli cloner implies that all coefficients $a_i$, with $i>0$ in the program state $\psi$ should be equal for the symmetric cloner.
The normalization condition for the program state $\psi$ then is given by, for any $i>0$:
\begin{equation}
a_0^2 + (d^2-1) a_i^2 = 1 , i>0,
\end{equation}
and according to Equation~\eqref{eq:generalized_f_b}, Bob's fidelity in each basis is given by the sum of $a_0^2$ and $\frac{(d^2-1)}{d+1}=d-1$ of the other $a^2_i$, which all take the same value for the UQCM. So we have
\begin{equation}
\frac{d+3}{2(d+1)} = \frac{a_0^2 + (d^2-1) a_i^2}{d+1},
\end{equation}
which implies
\begin{eqnarray}
a_0 &=& \sqrt{\frac{d+1}{2d}},\\
a_i &=& \sqrt{\frac{1}{2 d (d+1)}} , i>0.
\end{eqnarray}
This choice for $a_i$ implements the symmetric UQCM with the circuit in Figure~\ref{NqubitNGcircuit} for any number of qubits.

\section*{Conclusion and Outlook}
We have shown that the single-qubit Niu--Griffiths circuit~\cite{niugriffiths} can be generalized to an $N$-qubit Pauli cloner. We provided detailed analysis and examples for the $N=2$ case, alongside the analytical construction of both known and novel program states for $N=1$. Furthermore, we identified the specific program state that implements the symmetric $N$-qubit UQCM using the circuit shown in Figure~\ref{NqubitNGcircuit} for any number $N$ of qubits.\\
We have outlined the relationship between MUBs and Pauli errors in the general case, notably that Pauli cloners inherently introduce biases across the different MUBs, wheres the fidelity remains identical for all states within a specific MUB.\\
Furthermore, we have characterized how the program state $\ket{\psi}$ controls the disturbance of the Pauli cloner. In practical terms, this allows an eavesdropper to tune error rates for each type of Pauli error detected by the legitimate receiver. Such a capability can be used to obfuscate the unavoidable disturbance caused by any cloning device.\\
The structural properties of Pauli cloners, as well as extensive experimentation for small $N$ (using quantum machine learning and numerical techniques) leads us to conjecture that for any $N$-qubit Pauli channel, the optimal cloner (in terms of single-qubit fidelities, but presumably also by other measures) can be implemented with the circuit in Figure~\ref{NqubitNGcircuit}. A proof for the single-qubit case may be achievable by relaxing the assumption of equal basis fidelity found in the UQCM optimality proof~\cite{sixstate}.\\
Another open question remains: whether this construction generalizes to systems of dimension $d \neq 2^N$, with qutrits being the most immediate example. This would require to define analogues of the Pauli group and Pauli operators for such dimensions. Candidates to try would be higher-spin matrices and the corresponding cloner would be suited to clone quantum channels affected by noise generated by those operators. If this approach is successful, Pauli cloners could be viewed as the lowest dimensional example of a more general class of \textit{spin cloners}.\\

\begin{acknowledgments}
Circuits diagrams were generated with the quantikz~\cite{quantikz} package.\\
We thank Zi Chua, Lucas Euler and Tobias Meng for their feedback on the manuscript.
\end{acknowledgments}

\newpage
\bibliographystyle{ieeetr}
\bibliography{reference}

\appendix
\onecolumngrid
\newpage
\section{Effect of the noise channel on the fidelities}
\label{appendix:noise_effect}
We model an individual attack on the six-state protocol with a quantum circuit. The communication channel is represented by a single qubit with the two parties Alice and Bob at each end. Eve has access to an ancillary system and can interact with the communication line.

Alice prepares an initial state $\rho_A = \ket{\psi}\bra{\psi}$ which is transmitted to Bob through the communication line. Eve initializes her ancillary qubits and applies a unitary $U$ on the whole system:

\begin{equation}
\rho_{BE} = U(\rho_A\otimes \ket{0_E}\bra{0_E})U^\dagger
\end{equation}

Bob receives the reduced state $\rho_B = \operatorname{Tr}_E [\rho_{BE}]$ and the quantum fidelity between Alice and Bob can be written:

\begin{equation}
F_{AB,\ket{\psi}} = \bra{\psi} \rho_B \ket{\psi} 
\end{equation}

Now we consider that an error can occur between Alice's state preparation and Eve's unitary. We use the error model:

\begin{equation}
\mathcal{E}(\rho)
    = (1 - p_X - p_Y - p_Z) \, \rho
    + p_X \, X \rho X^\dagger 
    + p_Y \, Y\rho Y^\dagger 
    + p_Z \, Z \rho Z^\dagger
\end{equation}

To derive the relationship between the noisy fidelity $\tilde{F}_{AB}$ and the noiseless fidelity $F_{AB}$, we consider the specific case where Alice transmits the state $\ket{0}$. The evolution of the state under the noise model is given by:

\begin{align*}
\mathcal{E}(\ket{0}\bra{0}) 
&= (1 - p_X - p_Y - p_Z) \, \ket{0}\bra{0} 
    + p_X \, X \ket{0}\bra{0} X^\dagger 
    + p_Y \, Y \ket{0}\bra{0} Y^\dagger 
    + p_Z \, Z \ket{0}\bra{0} Z^\dagger \\
&= (1 - p_X - p_Y - p_Z) \, \ket{0}\bra{0} 
    + p_X \, \ket{1}\bra{1}
    + p_Y \, \ket{1}\bra{1} 
    + p_Z \, \ket{0}\bra{0} \\
&= (1 - p_X - p_Y) \, \ket{0}\bra{0} 
    + (p_X + p_Y) \, \ket{1}\bra{1}
\end{align*}

Consequently, the state Bob receives after Eve's unitary becomes:
\begin{align*}
\tilde{\rho}_{B, \ket{0}}
&= \operatorname{Tr}_E \left[ U \left( \mathcal{E}(\ket{0}\bra{0}) \otimes \ket{0_E}\bra{0_E} \right) U^\dagger \right] \\
&= (1 - p_X - p_Y) \, \operatorname{Tr}_E \left[ U \left( \ket{0}\bra{0} \otimes \ket{0_E}\bra{0_E} \right) U^\dagger \right]
    + (p_X + p_Y) \, \operatorname{Tr}_E \left[ U \left( \ket{1}\bra{1} \otimes \ket{0_E}\bra{0_E} \right) U^\dagger \right] \\
&= (1 - p_X - p_Y) \, \rho_{B, \ket{0}}
    + (p_X + p_Y) \, \rho_{B, \ket{1}}
\end{align*}

Here, $\rho_{B, \ket{0}}$ and $\rho_{B, \ket{1}}$ denote the states received by Bob in the noiseless case when Alice transmits states $\ket{0}$ and $\ket{1}$ respectively. Bob's noisy fidelity for the state $\ket{0}$ is given by:

\begin{align*}
\tilde{F}_{AB, \ket{0}} 
&= \bra{0} \tilde{\rho}_{B, \ket{0}} \ket{0} \\
&= (1 - p_X - p_Y) \, \bra{0} \rho_{B, \ket{0}} \ket{0}
    + (p_X + p_Y) \, \bra{0} \rho_{B, \ket{1}} \ket{0} \\
&= (1 - p_X - p_Y) \, \bra{0} \rho_{B, \ket{0}} \ket{0}
    + (p_X + p_Y) \left(1 - \bra{1} \rho_{B, \ket{1}} \ket{1} \right)\\
&= (1 - p_X - p_Y) \, F_{AB, \ket{0}}
    + (p_X + p_Y) \left( 1 - F_{AB, \ket{1}} \right) \\
&= (1 - p_X - p_Y) \, F_{AB, \ket{0}}
    - (p_X + p_Y) \, F_{AB, \ket{1}}
    + p_X + p_Y
\end{align*}

In the special case where $F_{AB, \ket{0}} = F_{AB, \ket{1}}$ --as is the case for the one-qubit Niu--Griffiths cloner-- the expression simplifies to:

\begin{equation}
\tilde{F}_{AB, \ket{0}} = (1 - 2p_X - 2p_Y)F_{AB, \ket{0}} + p_X + p_Y
\end{equation}

By the same reasoning, we find for the specific case when Alice transmits the $\ket{1}$ state:

\begin{equation}
\tilde{F}_{AB, \ket{1}} = (1 - 2p_X - 2p_Y)F_{AB, \ket{1}} + p_X + p_Y
\end{equation}

Together, these two expressions yield Equation~\eqref{eq:NGnoisy_fabz}, which describes how the average fidelity of the $Z$ basis $F_{AB,Z} = \frac{1}{2} \left( F_{AB,\ket{0}} + F_{AB, \ket{1}} \right)$ is affected by the error channel. The same reasoning applied to states $\ket{+}$, $\ket{-}$, $\ket{+i}$ and $\ket{-i}$ leads to Equations~\eqref{eq:NGnoisy_fabx} and~\eqref{eq:NGnoisy_faby}.

\section{QML approach}
\label{appendix:qcl}
To identify optimal cloners for the B92 states and the two-qubit twenty-states, we use the QML-based framework detailed in~\cite{qkdasQML}. Originally designed for optimizing attacks on QKD protocols, we explain here how it can be adapted to optimize cloners for a given set of states. 

\subsection{B92 States}

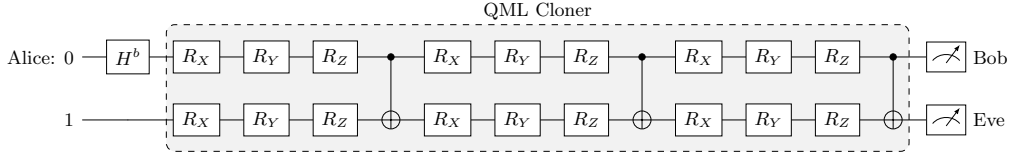
\begin{figure}[!h]
\begin{center}
\begin{tikzpicture}
\node[scale=0.8]{
\begin{quantikz}[thin lines, classical gap=0.2cm, column sep = 0.4cm]
\lstick{Alice: 0} & \gate{H^b} & \gate{R_X} \gategroup[2,steps=12,style={dashed,rounded corners,fill=black!5, inner xsep=0pt},background]{QML Cloner} & \gate{R_Y} & \gate{R_Z} & \ctrl{1} & \gate{R_X} & \gate{R_Y} & \gate{R_Z} & \ctrl{1} & \gate{R_X} & \gate{R_Y} & \gate{R_Z} & \ctrl{1} & \meter{} \rstick{Bob}\\
\lstick{1} & & \gate{R_X} & \gate{R_Y} & \gate{R_Z} & \targ{} & \gate{R_X} & \gate{R_Y} & \gate{R_Z} & \targ{} & \gate{R_X} & \gate{R_Y} & \gate{R_Z} & \targ{} & \meter{} \rstick{Eve}
\end{quantikz}
};
\end{tikzpicture}
\end{center}
\caption{Circuit implementation for optimizing cloning of the B92 states with QML. The 18 rotation gates in the ansatz are each associated with a trainable parameter.}
\label{fig:b92qcl}
\end{figure}

To optimize the cloning of B92 states, we use a two-qubit circuit. The first qubit encodes the state to be cloned. Alice randomly prepares it in states $\ket{0}$ or $\ket{+}$ using a Hadamard gate. The second qubit acts as an ancilla for the cloning process. A parametrized unitary operation $U(\Theta)$ is applied to the two qubits, yielding two approximate copies of the input state: one on qubit $0$ (Bob) and the other on qubit $1$ (Eve). The unitary is implemented using a QML ansatz containing $18$ trainable parameters, as illustrated in Figure~\ref{fig:b92qcl}.

Bob's fidelity when Alice transmits state $\ket{0}$ is defined as:

\begin{equation}
    F_{AB,\ket{0}}(\Theta) = \bra{0} \operatorname{Tr}_E \left[ U \ket{00}\bra{00} U^\dagger \right] \ket{0}
\end{equation}

Similarly, we define Bob's fidelity $F_{AB, \ket{+}}$ and Eve's fidelities $F_{AE, \ket{0}}$ and $F_{AE, \ket{+}}$. The average fidelities for Alice and Bob are then given by

\begin{align}
F_{AB} &= \left( F_{AB,\ket{0}} + F_{AB,\ket{+}}\right) / 2\\
F_{AE} &= \left( F_{AE,\ket{0}} + F_{AE,\ket{+}}\right) / 2
\end{align}

For a fixed target fidelity $f$ for Alice, we define the following loss function to maximize Eve's fidelity:

\begin{equation}
    \mathcal{L}(\Theta) = 10*\left( F_{AB}(\Theta) - f\right)^2 - F_{AE}(\Theta)
\end{equation}

This loss function is minimized using the Adam optimizer with different values of $f \in [0.5, 1]$ and a learning rate of $0.1$ over $100$ training steps. The fidelities obtained at the final step are displayed with blue diamonds in Figure~\ref{b92}.

\subsection{Program states for $N=2$}\label{twoqubitstateprep}

For this scenario, the approach is slightly different. We aim to study the Pauli cloner or the QID, but unlike the two-qubit case, performing a gridsearch over the 16 amplitudes of the 4-qubit program state $\ket{\psi}$ (as in Figure~\ref{fig:2qbcircuit}) is computationally impractical. Instead, we initialize the state using a parametrized circuit.

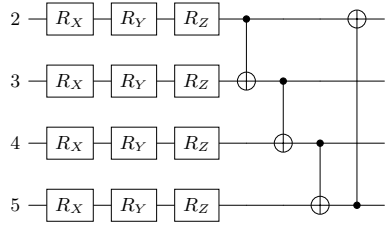
\begin{figure}[!h]
\begin{center}
\begin{tikzpicture}
\node[scale=0.8]{
\begin{quantikz}[thin lines, classical gap=0.2cm, column sep = 0.3cm]
\lstick{2} & \gate{R_X} & \gate{R_Y} & \gate{R_Z} & \ctrl{1} & & & \targ{} &\\
\lstick{3} & \gate{R_X} & \gate{R_Y} & \gate{R_Z} & \targ{} & \ctrl{1} & & &\\
\lstick{4} & \gate{R_X} & \gate{R_Y} & \gate{R_Z} & & \targ{} & \ctrl{1} & &\\
\lstick{5} & \gate{R_X} & \gate{R_Y} & \gate{R_Z} & & & \targ{} & \ctrl{-3} &\\
\end{quantikz}
};
\end{tikzpicture}
\end{center}
\caption{One of the $5$ layers used to initialize the software state $\ket{\psi}$ with a QML ansatz.}
\label{fig:healayer}
\end{figure}

Our ansatz consists of $5$ consecutive layers as shown in Figure~\ref{fig:healayer}, resulting in a total of $60$ trainable parameters. The cost function is the same as for the B92 states with $F_{AB}$ and $F_{AE}$ defined as the average fidelities over the five different bases. We perform several training rounds for different target fidelities $f \in [0.25, 1]$. The state preparation circuit is optimized using the Adam optimizer over 100 training steps and a learning rate of $0.1$.

\section{Analytical fidelities for the two-qubit Niu--Griffiths cloner on the twenty-state protocol}\label{2qubitNGfidelities}
For a general 4-qubits state $\ket{\psi} = a_0 \ket{0000} + \cdot + a_{15} \ket{1111}$ as input of the two-qubit Niu--Griffiths cloner, the fidelities for the twenty-state protocol can be written for each basis in the case the amplitudes $a_i$ are real:

\subsection{Basis $M_0$}
\begin{align}
F_{AB,M_0} &= a_0^2 + a_4^2 + a_{8}^2 + a_{12}^2\\
F_{AE,M_0} &= \frac{1}{4} + \frac{1}{2} ( a_0a_4 + a_1a_5 + a_2a_6 + a_3a_7 + a_4a_8 + a_5a_9 + a_6a_{10} + a_7a_{11} \nonumber \\
&\quad + a_8a_{12} + a_9a_{13} + a_{10}a_{14} + a_{11}a_{15} + a_0a_8 + a_1a_9 + a_2a_{10} + a_3a_{11} \nonumber \\
&\quad + a_4a_{12} + a_5a_{13} + a_6a_{14} + a_7a_{15} + a_{0}a_{12} + a_1a_{13} + a_2a_{14} + a_3a_{15})
\end{align}

\subsection{Basis $M_1$}
\begin{align}
F_{AB,M_0} &= a_0^2 + a_1^2 + a_{2}^2 + a_{3}^2\\
F_{AE,M_0} &= \frac{1}{4} + \frac{1}{2} ( a_0a_1 + a_0a_2 + a_0a_3 + a_1a_2 + a_1a_3 + a_2a_3 + a_4a_5 + a_4a_6 \nonumber \\
&\quad + a_4a_7 + a_5a_6 + a_5a_7 + a_6a_7 + a_8a_9 + a_8a_{10} + a_8a_{11} + a_9a_{10} \nonumber \\
&\quad + a_9a_{11} + a_{10}a_{11} + a_{12}a_{13} + a_{12}a_{14} + a_{12}a_{15} + a_{13}a_{14} + a_{13}a_{15} + a_{14}a_{15} )
\end{align}

\subsection{Basis $M_2$}
\begin{align}
F_{AB,M_0} &= a_0^2 + a_7^2 + a_{9}^2 + a_{14}^2\\
F_{AE,M_0} &= \frac{1}{4} + \frac{1}{2} ( a_0a_7 + a_0a_9 + a_0a_{14} + a_1a_6 + a_1a_8 + a_1a_{15} + a_2a_5 + a_2a_{11} \nonumber \\
&\quad + a_2a_{12} + a_3a_4 + a_3a_{10} + a_3a_{13} + a_4a_{10} + a_4a_{13} + a_5a_{11} + a_5a_{12} \nonumber \\
&\quad + a_6a_8 + a_6a_{15} a_7a_9 + a_7a_{14} + a_8a_{15} + a_9a_{14} + a_{10}a_{13} + a_{11}a_{12} )
\end{align}

\subsection{Basis $M_3$}
\begin{align}
F_{AB,M_0} &= a_0^2 + a_5^2 + a_{10}^2 + a_{15}^2\\
F_{AE,M_0} &= \frac{1}{4} + \frac{1}{2} ( a_0a_5 + a_0a_{10} + a_0a_{15} + a_1a_4 + a_1a_{11} + a_1a_{14} + a_2a_7 + a_2a_8 \nonumber \\
&\quad + a_2a_{13} + a_3a_6 + a_3a_9 + a_3a_{12} + a_4a_{11} + a_4a_{14} + a_5a_{10} + a_5a_{15} \nonumber \\
&\quad + a_6a_9 + a_6a_{12} + a_7a_8 + a_7a_{13} +a_8a_{13} + a_9a_{12} + a_{10}a_{15} + a_{11}a_{14} )
\end{align}

\subsection{Basis $M_4$}
\begin{align}
F_{AB,M_0} &= a_0^2 + a_6^2 + a_{11}^2 + a_{13}^2\\
F_{AE,M_0} &= \frac{1}{4} + \frac{1}{2} ( a_0a_6 + a_0a_{11} + a_0a_{13} + a_1a_7 + a_1a_{10} + a_1a_{12} + a_2a_4 + a_2a_9 \nonumber \\
&\quad + a_2a_{15} + a_3a_5 + a_3a_8 + a_3a_{14} + a_4a_9 + a_4a_{15} + a_5a_8 + a_5a_{14} \nonumber \\
&\quad + a_6a_{11} + a_6a_{13} + a_7a_{10} + a_7a_{12} + a_8a_{14} + a_9a_{15} + a_{10}a_{12} + a_{11}a_{13} )
\end{align}

\section{The QID for qubits: single-qubit one-to-two cloning }\label{QIDsinglequbit}
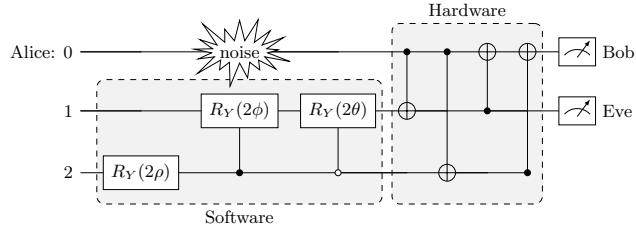
\begin{figure}[h!]
\centering
\begin{tikzpicture}
\node[scale=0.75]{
\tikzset{
noisy/.style={starburst,fill=white,draw=black,line
width=0.5pt,inner xsep=-4pt,inner ysep=-3pt}
}
\begin{quantikz}[thin lines, classical gap=0.09cm, column sep = 0.4cm, color=black,background color=white]
\lstick{Alice: 0}&\qw &\gate[1,style={noisy},label style=black]{\text{noise}} &\qw &\ctrl{1} \gategroup[3,steps=4,style={dashed,rounded corners,fill=black!5, inner xsep=0pt},background]{Hardware} &\ctrl{2}&\targ{}&\targ{}&\meter{} \rstick{Bob} \\
\lstick{1}&\qw \gategroup[2,steps=3,style={dashed,rounded corners,fill=black!5, inner xsep=0pt}, background, label style={label
position=below,anchor=north,yshift=-0.2cm}]{Software} &\gate{R_Y(2 \phi)}&\gate{R_Y(2 \theta)}&\targ{}&\qw &\ctrl{-1}&\qw &\meter{} \rstick{Eve} \\
\lstick{2}&\gate{R_Y(2 \rho)}&\ctrl{-1}&\ctrl[open]{-1}&\qw &\targ{}&\qw &\ctrl{-2}
\end{quantikz}
};
\end{tikzpicture}
    \caption{Quantum circuit implementing the QID for two qubits. Qubit 0 serves as both input and first output, while qubit 1 is the second output. Qubit 0 can be affected by Pauli noise, but only until Eve's controlled gates begin to act. The state initialization chosen here allows the most general real programs for the QID. Note the factors of 2 in front of the angles. Modifications to introduce relative complex phases have been left out here to declutter the diagram.}
    \label{1qubitQIDcircuit}
\end{figure}

The same analysis we did for the single-qubit NG cloner in Section~\ref{1qubitNG} can be performed for the single-qubit QID, with the quantum circuit given in Figure~\ref{1qubitQIDcircuit}. We use the same parameterization for the state preparation as for the NG cloner.

The first quantum circuit implementation of the QID for single qubits was published in~\cite{qid1997}, before the implementation of the asymmetric UQCM had been found. Our choice for the state preparation -- the three rotation gates in Figure~\ref{1qubitQIDcircuit} -- differ from that given in~\cite{qid1997} and leads to more convenient parameterizations later on.
The four CNOT gates in the circuit can be interpreted as the hardware of the QID, while the quantum state prepared on Eve's qubits 1 and 2 acts as the software. We denote the most general software state for the QID by
\begin{equation} \label{qidsoftware}
\ket{\psi} =
\begin{pmatrix}
a e^{i \alpha}\\ b e^{i \beta}\\ c e^{i \gamma}\\ d e^{i \delta}
\end{pmatrix}.
\end{equation}
with real numbers $a,b,c,d$ between 0 and 1.
Our goal is to relate the choice of $\ket{\psi}$ to the properties of the cloning machine it implements. Note that the circuit in Figure~\ref{1qubitQIDcircuit} produces only real states, that is $\alpha=\beta=\gamma=\delta=0$, just like in the case of the NG cloner. We will justify this in the following section.

\subsection{Analytical approach -- optimal QID programs for one qubit cloners}
\label{sec:1qbqid}
The circuit in Figure~\ref{1qubitQIDcircuit} implements the QID for a single qubit. The quantum state generated by the three $R_Y$-rotations should be thought of as the software for the QID, while the four CNOT gates should be thought of as the hardware.

We can view the circuit as a unitary and use that to calculate fidelities of the two output clones taken from qubits 0 and 1 with respect to the quantum state that Alice prepared. For this analysis, we consider the basis states of the $Z$, $X$ and $Y$ bases, namely
\[
\{\ket{0}, \ket{1}\}, \quad \{\ket{+}, \ket{-}\}, \quad \{\ket{+i}, \ket{-i}\}.
\]
The fidelity can then be computed for each state individually. For example, for state $\ket{0}$, the quantum fidelity with Bob's density matrix $\rho_B$ is given by
\begin{equation}
F_{AB, \ket{0}} = \bra{0} \rho_B \ket{0}.
\end{equation}
Inspection of the fidelity of all six MUB states of a single qubit in terms of the parameters in Equation~\eqref{qidsoftware} shows, that the 1-qubit QID clones states within the same basis with the same fidelity. This lets us define a single fidelity for each basis, for example $F_{AB, Z} = F_{AB, \ket{0}} = F_{AB, \ket{1}}$.
The fidelities can then be expressed directly as functions of the software state coefficients $a$, $b$, $c$ and $d$
as follows:

\begin{eqnarray}\label{1qubitfidelities}
F_{AB,Z} &=& a^2 + d^2 \label{fbz}\\
F_{AE,Z} &=& a^2 + b^2 \label{fez}\\
F_{AB,X} &=& ad \cos(\alpha-\delta) + bc \cos(\beta-\gamma) + \tfrac{1}{2} \label{fbx}\\
F_{AE,X} &=& ab \cos(\alpha-\beta)  + cd \cos(\gamma - \delta) + \tfrac{1}{2}\label{fex}\\
F_{AB,Y} &=& ad \cos(\alpha-\delta) - bc \cos(\beta-\gamma) + \tfrac{1}{2} \label{fby}\\
F_{AE,Y} &=& ab \cos(\alpha-\beta) - cd\cos(\gamma - \delta) + \tfrac{1}{2}\label{fey} 
\end{eqnarray}
Additionally, normalization of the software state gives:
\begin{equation}\label{QIDnorm}
a^2 + b^2 +c^2 +d^2 = 1
\end{equation}
These equations provide the basis for deriving the optimal QID program for a given cloning problem. To make precise what we mean by an optimal QID program, we state the cloning problem as the maximization of the weighted fidelities. For example, we can define the quality $Q$ of the cloner for Bob by $$ Q_B = x F_{ABX} + y F_{ABY} + z F_{ABZ}$$ for $x,y,z$ in the interval $[0,1]$, and similarly for $Q_E$ for Eve, where the same coefficients $x,y,z$ appear in Bob's and Eve's quality. We can now impose constraints, for example, a fixed value for $Q_B$. The optimization task then is to find $a,b,c,d$ and $\alpha,\beta,\gamma,\delta$ such that $Q_E$ is maximized. For example, this optimization approach with $x=y=z$ yields the UQCM, while for $x=y, z=0$  we find the PCCM in the $XY$ plane.\\
To justify that we only consider real-valued cloning software for the QID, we observe that for any $a,b,c,d$, the contribution of the first terms of $F_{AB,X}$ and  $F_{AB,Y}$ in Equations~\eqref{fbx} and~\eqref{fby} will always be maximized for $\alpha=\delta$. Similarly, the first terms in $F_{AE,X}$ and $F_{AE,Y}$ in~\eqref{fex} and~\eqref{fey} are maximized for $\alpha=\beta$. To maximize the remaining angle dependent terms, we note that if $x>y$, the weighted sum $Q_B = x F_{ABX} + y F_{ABY} + z F_{ABZ}$ is maximized for $\beta=\gamma$. Similarly, $Q_E = x F_{AEX} + y F_{AEY} + z F_{AEZ}$ is maximal for $\gamma=\delta$. This means that the average fidelity can be realized for $Q_B$ and $Q_E$ with the choice $\alpha=\beta=\gamma=\delta=0$, and we can choose the optimal software state $\psi$ to be real.\\
If $x<y$, we get maxima for the last two angle dependent terms for $\beta=\gamma-\pi$ and $\gamma=\delta+\pi$. So in this case, we can choose $\alpha=\beta=\delta=0$ and $\gamma=\pi$ to get the maximal average fidelities.\\
This shows that the optimal averaged fidelities can always be achieved with real-valued program states $\psi$. This justifies that our circuit in Figure~\ref{1qubitQIDcircuit} produces only real program states. From here on, to simplify the notation, we will treat $a,b,c,d$ as real (but not necessarily positive) numbers and ignore the angles $\alpha,\beta,\gamma,\delta$.

\subsection{One qubit QID programs}
\label{sec:1qb}

It is convenient to parameterize the QID software state $\ket{\psi}$ in terms of the three rotation angles used in the quantum circuit implementation in Figure~\ref{1qubitQIDcircuit}. In the computational basis, the column-vector representation of the state reads
\begin{equation} \label{QIDprogram}
\ket{\psi} =
\begin{pmatrix}
a \\ b \\ c \\ d
\end{pmatrix} =
\left(
\begin{array}{c}
\cos{\theta} \cos{\rho} \\[0.5mm]
\cos{\phi} \sin{\rho} \\[0.5mm]
\sin{\theta} \cos{\rho} \\[0.5mm]
\sin{\phi} \sin{\rho}
\end{array}
\right)
\end{equation}

A useful aspect of this parameterization is, that the parameters have intuitive interpretations. In particular, $\theta$ controls the ratio of the fidelities in the $X$ and the $Y$ bases, with $\theta=0$ corresponding to equal fidelities.
The parameters $\phi$ and $\rho$ parameterize the asymmetry between the two clones and the bias between the $Z$ and $X$ bases, but are independent of each other only in special cases. For the PCCM, for example, $\rho$ is fixed while $\phi$ controls the asymmetry between the two clones.

The optimal parameters of the QID for a given cloning task can then be determined systematically. One formulates the desired fidelity requirements in terms of the parameterization Equations~\eqref{1qubitfidelities} and solves the resulting system.\\
As examples, we now list some explicit program states for the QID.

\subsubsection*{Cloners with equal fidelity in the $X$ and $Y$ bases}
Because of the symmetries between $X$ and $Y$ in Equations~\eqref{fbx} to~\eqref{fey}, it is relatively straightforward to treat cloners with equal fidelities in these two bases analytically. The condition $\theta = 0$ ensures identical fidelities in the $X$ and $Y$ bases. In this case, the QID software state takes the form
\begin{equation}
\ket{\psi} =
\left(
\begin{array}{c}
\cos{\rho} \\[0.5mm]
 \cos(\phi) \sin(\rho) \\[0.5mm]
0 \\[0.5mm]
\sin(\phi) \sin(\rho)
\end{array}
\right)
\end{equation}
and the parameter $\rho$ controls the trade-off between fidelities in the $XY$-plane and in the $Z$-basis. The two extreme values $\rho = 0$ and $\rho = \pi$ yield the \textit{CNOT-cloner}, which clones the $Z$-basis perfectly
\[
F_{AB,Z} = F_{AE,Z} = 1,
\]
while outputting completely mixed states in the $X$ and $Y$ bases,
\[
F_{AB,X} = F_{AE,X} = F_{AB,Y} = F_{AE,Y} = \tfrac{1}{2}.
\]
Slightly more interesting is $\rho = \pi/2$, which provides an asymmetric cloner for the $Z$ basis with fidelities 
$$
F_{AB,Z} = \sin^2(\phi), \hspace{1cm} F_{AE,Z} = \cos^2(\phi)
$$
and
$$
F_{AB,X} = F_{AE,X} = F_{AB,Y} = F_{AE,Y} = \tfrac{1}{2}.
$$

As $\rho$ increases from $0$, the fidelity gradually shifts from the $Z$ basis to the $XY$-plane. The value $\rho = \pi/4$ maximizes this shift and yields the maximal fidelity in the $XY$-plane: it implements a PCCM with
\begin{align}
F_{AB,X} = F_{AB,Y} = \frac{1+\cos{\phi}}{2}\\
F_{AE,X} = F_{AE,Y} = \frac{1+\sin{\phi}}{2}
\end{align}
with $\phi$ controlling the asymmetry between the two clones. The symmetric PCCM is obtained for $\phi = \pi/4$.
Optimization for equal fidelities in all three bases with this symmetric condition leads to the UQCM. In particular, maximizing the sum $F_{AB,X} + F_{AB,Y} + F_{AB,Z}$ yields the condition $\rho = \arccos{\sqrt{2/3}}$ for which
\[
F_{AB} = F_{AE} = 5/6 .
\]
The asymmetric UQCM however cannot be obtained by simply varying $\phi$ at fixed $\rho$, but rather requires $\phi$ to satisfy
\begin{equation}
\phi = \arcsin\left(\frac{1}{2 \tan{\rho}} \pm \sqrt{\frac{1/2 - 3/4 \cos^2(\rho)}{\sin^2(\rho)})}\right) .
\end{equation}
Beyond the special cases, other combinations of $\rho$ and $\phi$ define a continuum of cloners with different trade-offs between fidelities in the $Z$-basis and $X$- and $Y$-bases. Although such cloners might seem of no use here, we will see in the next section how they can be applied in the presence of biased Pauli noise.\\
It is worth pointing out that, while the structure of the QID leads to fidelity equations with a symmetry between the $X$ and $Y$ basis, one can also find the cloners we discussed here, in the other basis pairs. So the QID can, for example, implement a PCCM in the $ZX$ plane, too -- but the calculations and solutions are more complicated.

\subsection{Optimal QID for cloning single qubits with Pauli noise}

\begin{figure}[!h]
\begin{center}
\includegraphics[width=0.5\columnwidth]{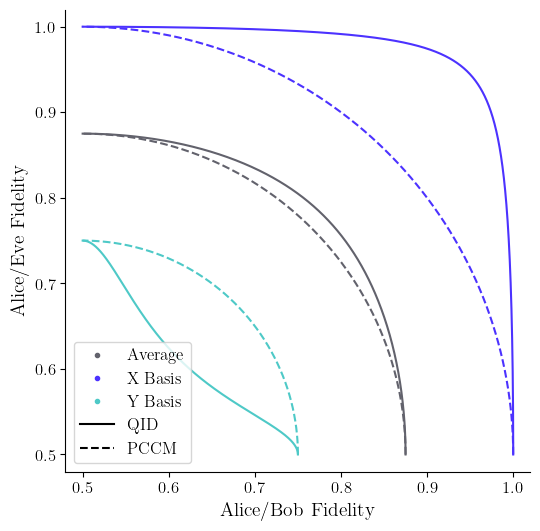}
\end{center}
\caption{PCCM and QID results for $p_X = 0.25$. The cloning performance is different for bases $X$ and $Y$. This enables the imbalanced cloner to outperform the UQCM on average.}
\label{fig:xynoise}
\end{figure}

In this section, we show that the QID can implement many more useful cloners than the ones we described in the last section, but we will take a different perspective. It was pointed out in~\cite{qkdasQML} that when errors affect the two bases of the BB84 protocol differently, so-called \textit{imbalanced cloners} can yield better average fidelities to Bob and Eve than the PCCM, which is the optimal attack on the protocol in the noiseless case. Here, we show that these imbalanced cloners can be reproduced by the QID with appropriate software. In addition, we extend this analysis to the three bases of the six-state protocol. Our perspective here becomes more application-oriented: if noisy qubits are to be cloned, what is the best QID program for the task? The answer to this question leads, depending on the noise model, to all the cloners mentioned above, and also new variants.\\
To introduce noise, we assume Alice and Bob communicate via a Pauli channel. The Pauli operators $$P_i \in \mathcal{P} = \{I,X,Y,Z\}$$ act on the qubit $\rho$ which Alice sends to Bob by $\sum_{i} p_i P_i \rho P_i$ , where the $p_i$ are the Pauli error rates and satisfy $\sum p_i = 1$. In particular, $p_I$ is the probability of no error occurring.

\begin{table}[h]
    \centering
    \begin{tabular}{c|c|c|c}
       & $X$-basis & $Y$-basis & $Z$-basis \\
       \hline
       $X$-error & 0 & 1 & 1\\
       $Y$-error & 1 & 0 & 1\\
       $Z$-error & 1 & 1 & 0
    \end{tabular}
    \caption{The effect of Pauli errors on the three bases. Value \textit{0} indicates that the basis is immune to the error (up to an overall phase), while \textit{1} indicates that the basis states are transformed by the error. The $X$-basis is immune to $X$-errors but susceptible to $Y$- and $Z$-errors, and similarly for the other two bases.}
    \label{tab:errors}
\end{table}
This model impacts the fidelities in the $X$, $Y$ and $Z$ bases differently. For clarity, Table~\ref{tab:errors} summarizes how each type of error affects the different bases. The resulting fidelities $\tilde{F}$ are given by:
\begin{align}
\tilde{F}_{AB,X} &= F_{AB,X}(1 - 2p_Y - 2p_Z) + p_Y + p_Z \label{eq:noisy_fabx} \\
\tilde{F}_{AB,Y} &= F_{AB,Y}(1 - 2p_X - 2p_Z) + p_X + p_Z \label{eq:noisy_faby} \\
\tilde{F}_{AB,Z} &= F_{AB,Z}(1 - 2p_X - 2p_Y) + p_X + p_Y \label{eq:noisy_fabz}
\end{align}
and similarly for the fidelities between Alice and Eve.
\subsubsection{XY-Bases imbalanced cloner}
For the sake of simplicity, we reproduce the imbalanced cloner of~\cite{qkdasQML} in the $XY$ bases instead of $ZX$. Because $Z$ errors affect bases $X$ and $Y$ the same way, we set $p_Z = 0$ and only consider $X$ and $Y$ errors. We identify the parameters $(\theta, \rho, \phi)$ which effectively produce an imbalanced cloner for given $p_X$ and $p_Y$ to be:
\begin{equation}
\theta = \arcsin\!\left( \sin\!\left( 2 \arctan\!\frac{p_X - p_Y}{1 - p_X - p_Y}\right) \sin(2\phi) \right)
\end{equation}
\begin{equation}
\rho = \pi/4
\end{equation}
Varying $\phi \in [0, \pi/2]$ changes the symmetry of the cloner, yielding a range of fidelities displayed in Figure~\ref{fig:xynoise} for an example with $p_X = 0.25$ and $p_Y = 0$. These results match exactly what was shown in~\cite{qkdasQML}. By strategically reducing fidelity in the affected $Y$-basis, the cloner gains more in the unaffected $X$-basis, allowing it to perform better than the PCCM on average. For $\phi = 0$, the asymmetry of the cloner advantages Eve, reducing Bob's fidelities to $F_{AB,X} = F_{AB,Y} = 1/2$ while Eve's fidelities are $F_{AB,X} = 1 - p_Y$ and $F_{AB,Y} = 1 - p_X$. Conversely, for $\phi = \pi/2$, it favors Bob with $F_{AE,X} = F_{AE,Y} = 1/2$, $F_{AE,X} = 1 - p_Y$ and $F_{AE,Y} = 1 - p_X$. For $\phi = \pi/4$:
\begin{equation}
    \theta = 2 \arctan \frac{p_X - p_Y}{1 - p_X - p_Y}
\end{equation}
gives the symmetric imbalanced cloner. When $p_X = p_Y$, $\theta = 0$ and we find the symmetric XY-PCCM described in Section~\ref{sec:1qb} as a special case.

\subsubsection{Generalized imbalanced cloners}

\begin{figure}[!h]
\begin{center}
\includegraphics[width=0.5\columnwidth]{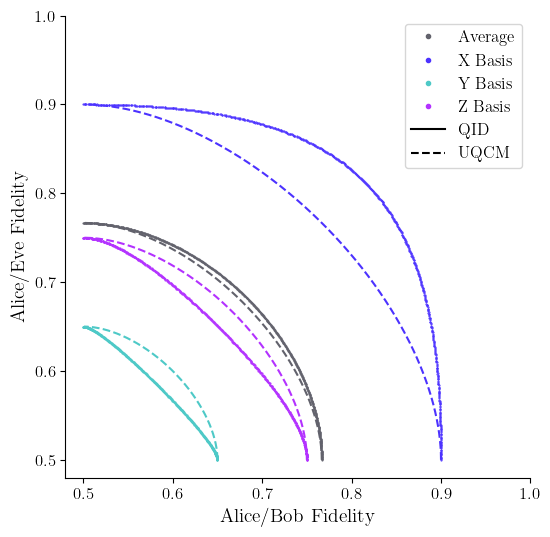}
\end{center}
\caption{UQCM and QID results for a communication channel with $p_X = 0.25$ and $p_Z = 0.1$. The cloning performance is different for the three bases of the six-state protocol. This enables the QID cloner to outperform the UQCM on average.}
\label{fig:sixstatenoise}
\end{figure}

We now extend this analysis to include the $Z$-basis, so that we now consider the three bases of the six-state protocol. As this set is maximal in terms of single-qubit mutually unbiased bases (MUBs), it provides the most general scenario for single-qubit cloning under our noise model.
In this broader setting, the additional complexity introduced by the third basis complicates an analytical analysis such as in the previous section. Instead, we rely on numerical methods to explore the achievable fidelities. Our results demonstrate that the known optimal attack for the six-state protocol, the UQCM, can be outperformed for specific values of $(p_X, p_Y, p_Z)$. More generally, the numerical evidence suggests that whenever a noise model $(p_X, p_Y, p_Z)$ introduces an imbalance between the three bases $X$, $Y$ and $Z$, it is possible to outperform the UQCM. An example for such a noise model is displayed in Figure~\ref{fig:sixstatenoise}.

This observation is in fact not particularly surprising. Consider, for instance a noise model with $p_Z = p_X = 0.25$ and $p_Y = 0$. In that case, the fidelities in the $Y$-basis are fixed at $F_{AB,Y} = F_{AE,Y} = 1/2$ regardless of the parameters $(\theta, \rho, \phi)$, while the fidelities in the $Z$ and $X$ bases are not fixed but affected equally. This situation is therefore equivalent to a noisy BB84 protocol without imbalance between the two bases. It is well known that in this setting, the PCCM is the optimal cloner and outperforms the UQCM. If we keep $p_Y = 0$ but now allow $p_Z \neq p_X$ with $p_Z + p_X = 0.5$, the problem reduces to an imbalanced version of BB84. As shown in the previous section, the corresponding imbalanced cloner then outperforms the PCCM.\\
We have demonstrated that in the single-qubit setting, the capabilities of the QID extend beyond known previous results. As a matter of fact, real valued QID programs, parametrized by $\theta, \rho, \phi$, can be found such that optimal average fidelity is achieved for any kind of imbalanced Pauli noise. The known (asymmetric) UQCM, PCCM and imbalanced cloners are all special cases of such biased cloners.
We expect these biased cloners (and as a special case, the imbalanced cloners of~\cite{qkdasQML}) to be optimal in the sense that higher average fidelities cannot be achieved by any unitary (we only analysed the QID based ones) for the noise models they are associated with. However, a formal proof of optimality is not provided here. A brief discussion on how this proof could be achieved is included in the conclusion of this paper.

\section{Fidelities for the two-qubit QID on the twenty-state protocol} \label{appendix:20fids}
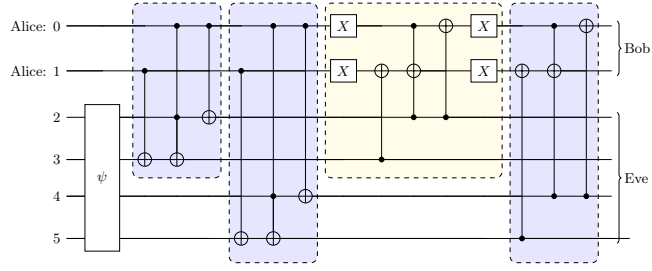
\begin{figure}[!h]
\begin{center}
\begin{tikzpicture}
\node[scale=0.6]{
\begin{quantikz}[thin lines, classical gap=0.2cm, column sep = 0.4cm]
\lstick{Alice: 0} & \qw & \qw \gategroup[4,steps=3,style={dashed,rounded corners,fill=blue!10, inner xsep=0pt},background]{} & \ctrl{3} & \ctrl{2} & \qw \gategroup[6,steps=3,style={dashed,rounded corners,fill=blue!10, inner xsep=0pt},background]{}& \ctrl{5} & \ctrl{4} & \gate{X} \gategroup[4,steps=5,style={dashed,rounded corners,fill=yellow!10, inner xsep=0pt},background]{} & \qw & \ctrl{1} & \targ{} & \gate{X} & \qw \gategroup[6,steps=3,style={dashed,rounded corners,fill=blue!10, inner xsep=0pt},background]{} & \ctrl{1} & \targ{} & \rstick[2]{Bob}\\
\lstick{Alice: 1} & \qw & \ctrl{2} & \qw & \qw & \ctrl{4} & \qw & \qw & \gate{X} & \targ{} & \targ{} & \qw & \gate{X} & \targ{} & \targ{} & \qw &\\
\lstick{2} & \gate[4]{\makebox[1.5em]{$\psi$}} & \qw & \ctrl{1} & \targ{} & \qw & \qw & \qw & \qw & \qw & \ctrl{-1} & \ctrl{-2} & \qw & \qw & \qw & \qw & \rstick[4]{Eve}\\
\lstick{3} & \qw & \targ{} & \targ{} & \qw & \qw & \qw & \qw & \qw & \ctrl{-2} & \qw & \qw & \qw & \qw & \qw & \qw &\\
\lstick{4} & \qw & \qw & \qw & \qw & \qw & \ctrl{1} & \targ{} & \qw & \qw & \qw & \qw & \qw & \qw & \ctrl{-3} & \ctrl{-4} &\\
\lstick{5} & \qw & \qw & \qw & \qw & \targ{} & \targ{} & \qw & \qw & \qw & \qw & \qw & \qw & \ctrl{-4} & \qw & \qw  & \qw &
\end{quantikz}
};
\end{tikzpicture}
\end{center}
\caption{A circuit implementation of the Quantum Information Distributor for two-qubit states. The box $\psi$ prepares the initial state on qubits 2 to 5, the program for the QID. The three purple blocks can be seen as the higher dimensional analogues of the three CNOT gates in the QID for one qubit, the yellow block is the analogue of the fourth CNOT, but differs in the shift direction from the others.}
\label{fig:2qbcircuit}
\end{figure}

A quantum circuit implementing the QID for cloning two-qubit states is given by
Our implementation of the QID now requires 4 ancillary qubits for a total of 6 qubits and is represented in Figure~\ref{fig:2qbcircuit}.

The sixteen-dimensional state $\ket{\psi}$ which is prepared on qubits 2 to 5 acts as software for the two-qubit QID. For example, the choice 
\begin{equation*}
\ket{\psi} = \tfrac{1}{10}(2,0,0,0,1,1,0,0,1,0,1,0,1,0,0,1)
\end{equation*}
implements the symmetric UQCM .\\

For a general 4-qubits state $\ket{\psi} = a_0 \ket{0000} + \cdot + a_{15} \ket{1111}$ as input of the two-qubit QID, the fidelities for the twenty-state protocol can be written for each basis in the case the amplitudes $a_i$ are real:

\subsection{Basis $M_0$}
\begin{align}
F_{AB,M_0} &= a_0^2 + a_5^2 + a_{10}^2 + a_{15}^2\\
F_{AE,M_0} &= a_0^2 + a_1^2 + a_2^2 + a_3^2 
\end{align}

\subsection{Basis $M_1$}

$M_1$ is the only basis for which the fidelities are not uniform across all four states, differing between the two state pairs $(s_0, s_2)$ and $(s_1, s_3)$:

\begin{enumerate}
\item For states $s_0$ and $s_2$:
\begin{align}
F_{AB,M_1,s_{0,2}} &= \frac{1}{4} + \frac{1}{2} (
a_0a_{10} + a_0a_{15} + a_0a_5 + a_1a_{11} + a_1a_{14} + a_1a_4 + a_{10}a_{15} + a_{10}a_5 \nonumber \\
&\quad + a_{11}a_{14} + a_{11}a_4 + a_{12}a_2 + a_{12}a_7 + a_{12}a_9 + a_{13}a_3 + a_{13}a_6 + a_{13}a_8 \nonumber \\
&\quad + a_{14}a_4 + a_{15}a_5 + a_2a_7 + a_2a_9 + a_3a_6 + a_3a_8 + a_6a_8 + a_7a_9) \\
F_{AE,M_1,s_{0,2}} &= \frac{1}{4} + \frac{1}{2} (
a_0a_{1} + a_0a_{2} + a_0a_3 + a_1a_{2} + a_1a_{3} + a_{10}a_{11} + a_{10}a_{8} + a_{10}a_9 \nonumber \\
&\quad + a_{11}a_{8} + a_{11}a_9 + a_{12}a_{13} + a_{12}a_{14} + a_{12}a_{15} + a_{13}a_{14} + a_{13}a_{15} + a_{14}a_{15} \nonumber \\
&\quad + a_{2}a_3 + a_4a_5 + a_4a_6 + a_4a_7 + a_5a_6 + a_5a_7 + a_6a_7 + a_8a_9)
\end{align}

\item For states $s_1$ and $s_3$:
\begin{align}
F_{AB,M_1,s_{1,3}} &= \frac{1}{4} + \frac{1}{2} (
a_0a_{10} + a_0a_{15} + a_0a_5 + a_1a_{11} + a_1a_{14} + a_1a_4 + a_{10}a_{15} + a_{10}a_5 \nonumber \\
&\quad + a_{11}a_{14} + a_{11}a_4 - a_{12}a_2 - a_{12}a_7 + a_{12}a_9 - a_{13}a_3 - a_{13}a_6 + a_{13}a_8 \nonumber \\
&\quad + a_{14}a_4 + a_{15}a_5 + a_2a_7 - a_2a_9 + a_3a_6 - a_3a_8 - a_6a_8 - a_7a_9 ) \\
F_{AE,M_1,s_{1,3}} &= \frac{1}{4} + \frac{1}{2} (
a_0a_{1} + a_0a_{2} + a_0a_3 + a_1a_{2} + a_1a_{3} + a_{10}a_{11} - a_{10}a_{8} - a_{10}a_9 \nonumber \\
&\quad - a_{11}a_{8} - a_{11}a_9 + a_{12}a_{13} - a_{12}a_{14} - a_{12}a_{15} - a_{13}a_{14} - a_{13}a_{15} + a_{14}a_{15} \nonumber \\
&\quad + a_{2}a_3 + a_4a_5 + a_4a_6 + a_4a_7 + a_5a_6 + a_5a_7 + a_6a_7 + a_8a_9 )
\end{align}
\end{enumerate}

\subsection{Basis $M_2$}

\begin{align}
F_{AB,M_2} &= \frac{1}{4} + \frac{1}{2} (
a_0a_{10} + a_0a_{15} + a_0a_5 - a_1a_{11} - a_1a_{14} + a_1a_4 + a_{10}a_{15} + a_{10}a_5 \nonumber \\
&\quad + a_{11}a_{14} - a_{11}a_4 - a_{12}a_9 - a_{13}a_8 - a_{14}a_4 + a_{15}a_5 - a_{2}a_7 - a_{3}a_6 ) \\
F_{AE,M_2} &= \frac{1}{4} + \frac{1}{2} (
a_0a_{1} + a_0a_{2} + a_0a_3 + a_1a_{2} + a_1a_{3} - a_{10}a_{11} - a_{12}a_{13} - a_{14}a_{15} \nonumber \\
&\quad + a_{2}a_{3} + a_{4}a_5 - a_{4}a_{6} - a_{4}a_{7} - a_{5}a_{6} - a_{5}a_{7} + a_{6}a_{7} - a_{8}a_{9} )
\end{align}

\subsection{Bases $M_3$ and $M_4$}

Both bases $M_3$ and $M_4$ have the same fidelities for Alice and Bob.

\begin{align}
F_{AB,M_{3,4}} &= \frac{1}{4} + \frac{1}{2} (
a_0a_{10} + a_0a_{15} + a_0a_5 - a_1a_{4} + a_{10}a_{15} + a_{10}a_5 - a_{11}a_{14} + a_{15}a_5 ) \\
F_{AE,M_{3, 4}} &= \frac{1}{4} + \frac{1}{2} (
a_0a_{1} + a_0a_{2} + a_0a_3 + a_1a_2 + a_1a_{3} + a_2a_{3} - a_{4}a_{5} - a_{6}a_{7})
\end{align}

\end{document}